\documentclass{article}
\usepackage[english]{babel}
\usepackage{here}
\usepackage{cite}
\usepackage[a4paper,top=2cm,bottom=2cm,left=3cm,right=3cm,marginparwidth=1.75cm]{geometry}

\usepackage{amsmath}
\usepackage{graphicx}
\usepackage{comment}
\usepackage{wrapfig}
\usepackage{caption}
\usepackage{subcaption}
\usepackage{amsmath}
\usepackage{amssymb}
\usepackage{authblk} 
\usepackage{booktabs} 

\usepackage{setspace}
\usepackage{lineno}

\usepackage[colorlinks=true, allcolors=blue]{hyperref}

\title{\textbf{High-Statistics Study of Antiproton Annihilation at Rest on Argon in a LArTPC}}

\author[1,$\dagger$]{M.~Tanaka}
\author[1,$\ddagger$]{K.~Yorita}
\author[2]{S.~Arai}
\author[2]{K.~Ishikawa}
\author[2]{Y.~Utsumi}
\author[2]{C.~Watanabe}
\author[2]{Y.~Yano}
\author[3]{T.~Akaishi}
\author[4]{K.~Aoyama}
\author[1]{S.~Arai}
\author[5]{T.~Aramaki}
\author[6,7]{T.~Hashimoto}
\author[8]{G.~Karagiorgi}
\author[1]{M.~Kimura}
\author[9]{Y.~Kimura}
\author[5]{J.~Leyva}
\author[6]{Y.~Ma}
\author[11]{J.~Mitchell}
\author[10]{T.~Nanamura}
\author[12]{H.~Odaka}
\author[11]{M.~Sasaki}
\author[6]{S.~Takashima}
\author[8]{S.~Tandon}
\author[12]{R.~Tatsumi}
\author[5]{A.~Barnett} 
\author[11]{N.~Cannady}
\author[5]{C.~Earley} 
\author[13]{K.~Sakai}
\author[5]{J.~Zeng} 

\affil[1]{Department of Physics, University of Tokyo, Tokyo 113-0033, Japan}
\affil[2]{Department of Physics, Waseda University, Tokyo 169-8555, Japan}
\affil[3]{Research Center for Nuclear Physics (RCNP), Osaka University, Suita, Osaka 567-0047, Japan}
\affil[4]{Japan Aerospace Exploration Agency (JAXA), Sagamihara, Kanagawa 252-5210, Japan}
\affil[5]{Department of Physics, Northeastern University, Boston, MA 02115, USA}
\affil[6]{RIKEN Nishina Center, Hirosawa 2-1, Wako-shi, Saitama 351-0198, Japan}
\affil[7]{RIKEN Pioneering Research Institute, Wako, Saitama 351-0198, Japan}
\affil[8]{Department of Physics, Columbia University, New York, NY 10027, USA}
\affil[9]{Research Center of Accelerator and Radioisotope Science (RARiS), Tohoku University, Sendai, Miyagi 982-0826, Japan}
\affil[10]{Department of Physics, Tohoku University, Sendai 980-8578, Japan}
\affil[11]{NASA Goddard Space Flight Center, Greenbelt, MD 20771, USA}
\affil[12]{Department of Earth and Space Science, Graduate School of Science, Osaka University, Osaka 560-0043, Japan}
\affil[13]{Enrico Fermi Institute, The University of Chicago, 5640 South Ellis Ave Chicago IL 60637, USA}

\date{\today}

\begin{document}

\maketitle

\footnotetext[1]{$\dagger$ masa@kylab.phys.s.u-tokyo.ac.jp}
\footnotetext[2]{$\ddagger$ kohei.yorita@phys.s.u-tokyo.ac.jp}

\begin{abstract}
This paper reports a high-statistics study of antiproton--argon ($\bar{p}\text{--}\text{Ar}$) annihilation at rest in the J-PARC T98 experiment using a single-phase Liquid Argon Time Projection Chamber (LArTPC). A dataset comprising over $10^{5}$ annihilation events was acquired together with comprehensive calibration data. In parallel, a dedicated Geant4-based simulation framework was developed to model the secondary beam transport and detector response. Together, these experimental and computational resources provide a foundation for benchmarking hadronic interaction models and for future LArTPC-based antimatter searches, such as GRAMS. As an initial application of this framework, we compare the measured annihilation topologies, localized energy profiles, and spatial cluster multiplicities with predictions from three hadronic interaction generators: Fritiof, INCL++, and GiBUU. While all models reproduce the overall features of the total energy deposition, GiBUU provides the most consistent empirical description of localized energy deposition near the annihilation vertex and spatial cluster multiplicities at larger radii, reflecting the key roles of secondary nucleon knockout and in-medium pion absorption. These results provide a high-statistics experimental benchmark for modeling low-energy antiproton interactions in liquid argon.
\end{abstract}

\clearpage

\section{Introduction}

Antiproton--nucleus ($\bar{p}\text{--}A$) annihilation at rest represents a fundamental process in hadronic physics, yielding striking multi-particle final states~\cite{Chamberlain:1955ns,Klempt:2005pp}. When a low-energy antiproton slows down and stops within a target medium, it is captured to form an exotic atom before ultimately annihilating with a nucleon near the nuclear surface. This interaction primarily produces a high-multiplicity shower of pions and recoiling nucleons, forming a distinct "annihilation star" topology. Liquid Argon Time Projection Chambers (LArTPCs) excel at capturing these complex three-dimensional topologies with high spatial and calorimetric resolution~\cite{Rubbia:1977zz}.

However, interpreting these interactions is complicated by intranuclear cascade dynamics. Secondary hadrons produced in the primary annihilation must traverse the parent nucleus ($^{40}\text{Ar}$), undergoing extensive Final State Interactions (FSI) such as rescattering, charge exchange, and absorption. These FSI processes significantly alter the kinematic distributions and particle multiplicities of the outgoing products. Consequently, precise modeling of FSI in heavy nuclei remains a critical necessity for current Monte Carlo hadronic generators.

Constraining hadronic transport models through experimental $\bar{p}\text{--}\text{Ar}$ annihilation data has broad implications for both cosmic-ray physics and fundamental interactions. In cosmic-ray physics, searches for exotic antimatter signatures (such as low-energy antiprotons, antideuterons, and antihelium nuclei) serve as promising indirect probes for dark matter~\cite{Cholis:2020twh}. While current experiments like GAPS utilize semiconductor detectors to achieve particle identification via atomic X-rays \cite{Mori:2002}, the next-generation GRAMS experiment adopts LArTPC technology to achieve even higher sensitivity by precisely reconstructing the complex hadronic tracks forming the characteristic annihilation star~\cite{Aramaki:2019bpi}. Furthermore, $\bar{p}\text{--}\text{Ar}$ annihilation modeling provides vital inputs for constraining FSI uncertainties in rare-event searches, including baryon-number-violating neutron--antineutron ($n\text{--}\bar{n}$) oscillations evaluated in large LArTPCs like DUNE \cite{Mohapatra:2009wp,DUNE:2020ypp}. Therefore, high-statistics benchmarking of antiproton interactions in liquid argon is essential to refine these nuclear models and reduce systematic uncertainties across future LArTPC-based physics programs.

In light of these physics motivations, initial experimental progress was made by the LArIAT experiment at Fermilab, which achieved the first observation of antiproton annihilation at rest on argon nuclei using a LArTPC \cite{LArIAT:2024lbx}. Because the LArIAT test beamline extracted the secondary beam without particle separation, the incoming beam was dominated by a large background of pions and other particles. Consequently, the analysis yielded a modest sample of candidate $\bar{p}$ annihilation events identified through manual hand-scanning and automated tracking algorithms, serving as a foundational proof-of-concept for $\bar{p}\text{--}\text{Ar}$ interactions.

While LArIAT demonstrated the feasibility of measuring $\bar{p}\text{--}\text{Ar}$ annihilation in LAr, high-statistics data are indispensable for rigorously constraining intranuclear cascade models and hadronic transport inside heavy argon nuclei ($^{40}\text{Ar}$). The J-PARC T98 experiment was designed to address this requirement by deploying a prototype LArTPC at the K1.8BR secondary beamline in the J-PARC Hadron Hall \cite{Agari:2012_K18BR}. Taking advantage of an electrostatic separator (ES) to effectively sweep out pion contaminants, T98 successfully collected an unprecedented high-statistics dataset of $>10^5$ antiproton--argon reaction events at $0.7~\text{GeV}/c$ in February 2025.

In this paper, we construct a Monte Carlo simulation framework calibrated and validated against reference datasets (such as cosmic muons and proton beam controls) to accurately reproduce the detector response. Utilizing this established framework, we investigate $0.7~\text{GeV}/c$ $\bar{p}$ annihilation events—where theoretical uncertainties in hadronic interaction models remain significant—by benchmarking three prominent hadronic models: two implemented within Geant4—the Fritiof model \cite{Andersson:1987_FTF, Agostinelli:2003_Geant4} and the Liège Intranuclear Cascade model (INCL++) \cite{Mancusi:2014_INCL, Zharenov:2023_INCL_pbar}—alongside the independent GiBUU transport model \cite{Buss:2012_GiBUU}.

The remainder of this paper is composed
of five additional sections.
Section~\ref{sec:setup} describes the experimental configuration of the J-PARC T98 and the LArTPC detector. Section~\ref{sec:reconstruction} details the 3D track and vertex reconstruction procedures. Section~\ref{sec:simulation} introduces the implementations of the simulation framework and the hadronic models. Section~\ref{sec:results} presents the results and discussion based on Data-MC comparisons. Finally, Section~\ref{sec:conclusion} summarizes our findings and outlines future prospects.

\section{Experimental Setup and Beam Exposure (J-PARC T98)}
\label{sec:setup}

\subsection{The K1.8BR Beamline and Mass Separation}
\label{sec:beam_sim}
\begin{figure}[htbp]
  \centering
  \includegraphics[width=0.95\linewidth]{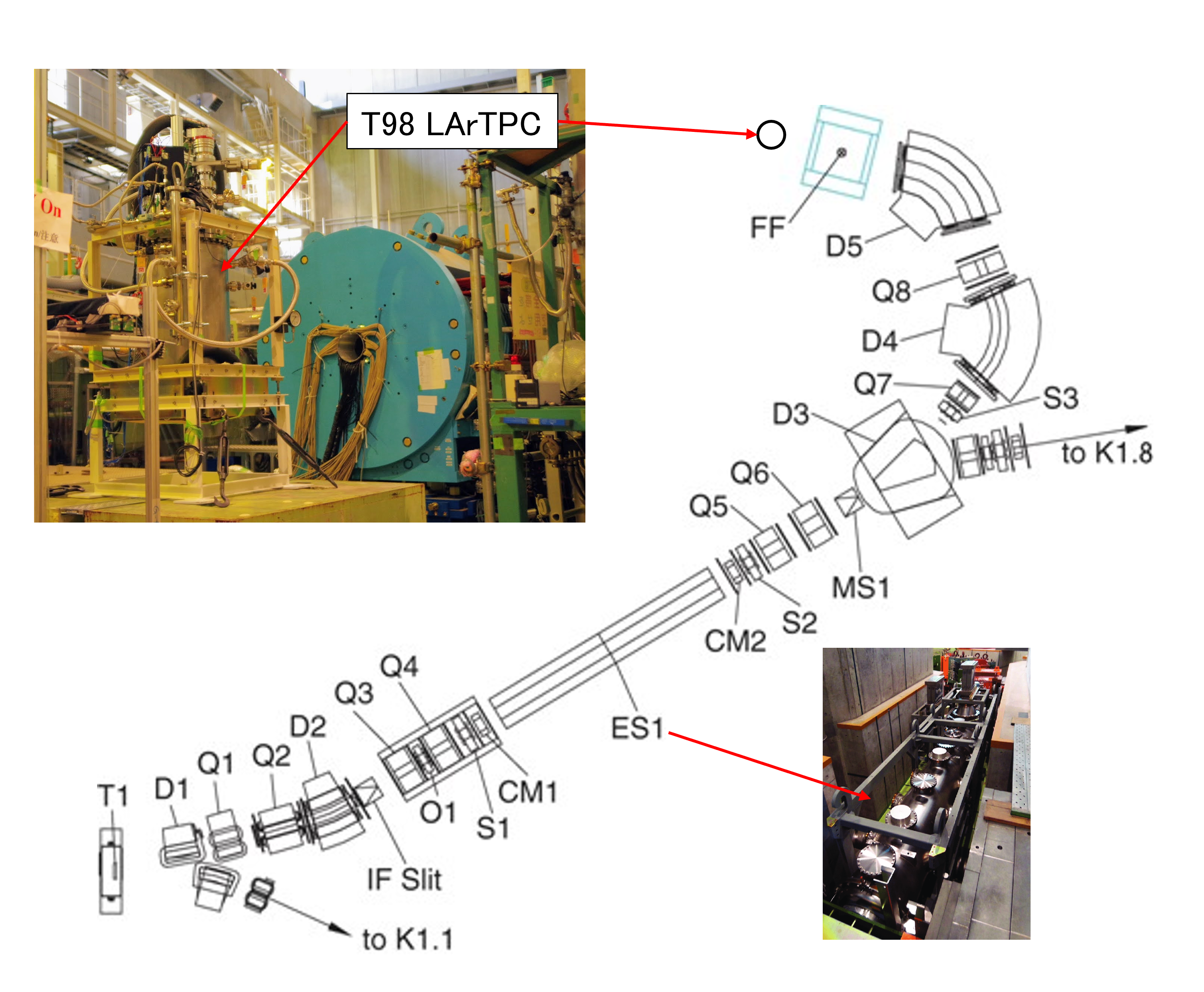}
  \caption{Schematic layout of the K1.8BR secondary beamline in the J-PARC Hadron Hall, accompanied by photographs of the Electrostatic Separator (ES1, bottom right) and the T98 LArTPC detector installed at the final focus area (FF, top left). }
  \label{fig:k18br_beamline}
\end{figure}

The J-PARC T98 experiment was conducted at the K1.8BR secondary beamline located in the J-PARC Hadron Hall \cite{Agari:2012_K18BR}. Figure~\ref{fig:k18br_beamline} shows the schematic layout of the K1.8BR beamline together with photographs of the Electrostatic Separator (ES1) and the installed T98 LArTPC. Secondary charged particles with a momentum up to $1.2~\text{GeV}/c$ are extracted by colliding the $30~\text{GeV}$ primary proton beam onto a gold production target. In the production stage, the secondary negative beam is overwhelmingly dominated by pions ($\pi^-$), producing an extremely low initial antiproton-to-pion ratio ($\bar{p}/\pi^-$) of approximately $0.1\%$ \cite{Agari:2012_K18BR, Sanford:1967_SW}. As indicated in Fig.~\ref{fig:k18br_beamline}, the T98 LArTPC detector was installed $\sim$ 2.5m  downstream of the final focus point (FF) located at the most downstream end of the beamline.

\begin{figure}[htbp]
  \centering
  \includegraphics[width=0.95\linewidth]{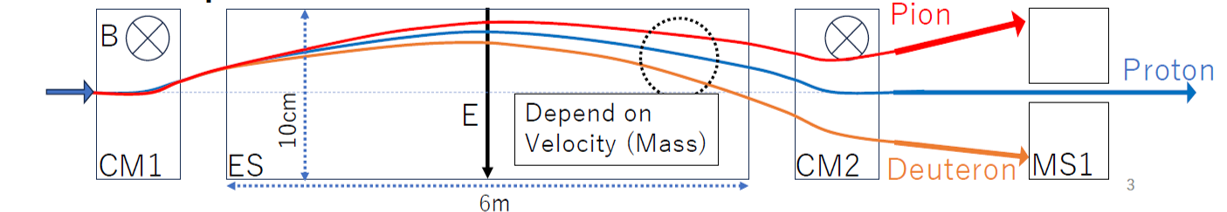}
    \caption{Schematic diagram of the mass separation principle. Particles with equal momentum but different masses (and thus different velocities $v$) experience an electrostatic force ($F = qE$) in the vertical electric field ($E$) within the 6-m Electrostatic Separator (ES), resulting in different transit times ($t = L/v$) and transverse deflections before reaching the mass slit (MS1).}
  \label{fig:mass_separation}
\end{figure}

To deliver a high-purity antiproton beam to the LArTPC, the K1.8BR beamline incorporates a mass separation system consisting of correction magnets (CM1, CM2) and a 6-m-long Electrostatic Separator (ES) \cite{Ieiri:2008zzc, Agari:2012_K18BR}, as illustrated by the schematic in Fig.~\ref{fig:mass_separation}. Particles entering the ES experience a vertical electric field ($E$), which exerts an electrostatic force ($F = qE$). Because particles with identical momentum possess different velocities ($v$) depending on their masses, their transit times through the separator ($t = L/v$) and resulting transverse deflections diverge accordingly. The vertical slit (MS1) placed downstream of CM2 physically blocks the unwanted pion contamination while selectively passing antiprotons.

Through this mass separation system, the $\bar{p}/\pi^-$ ratio is significantly enriched from less than $0.1\%$ to roughly $1:1$ at the beamline focus, effectively delivering a high-density antiproton beam suitable for downstream particle identification and irradiation into the LArTPC.

\subsection{The LArTPC Detector and Readout System}
\label{sec:lartpc_detector}

\begin{figure}[htbp]
  \centering
  \includegraphics[width=0.95\linewidth]{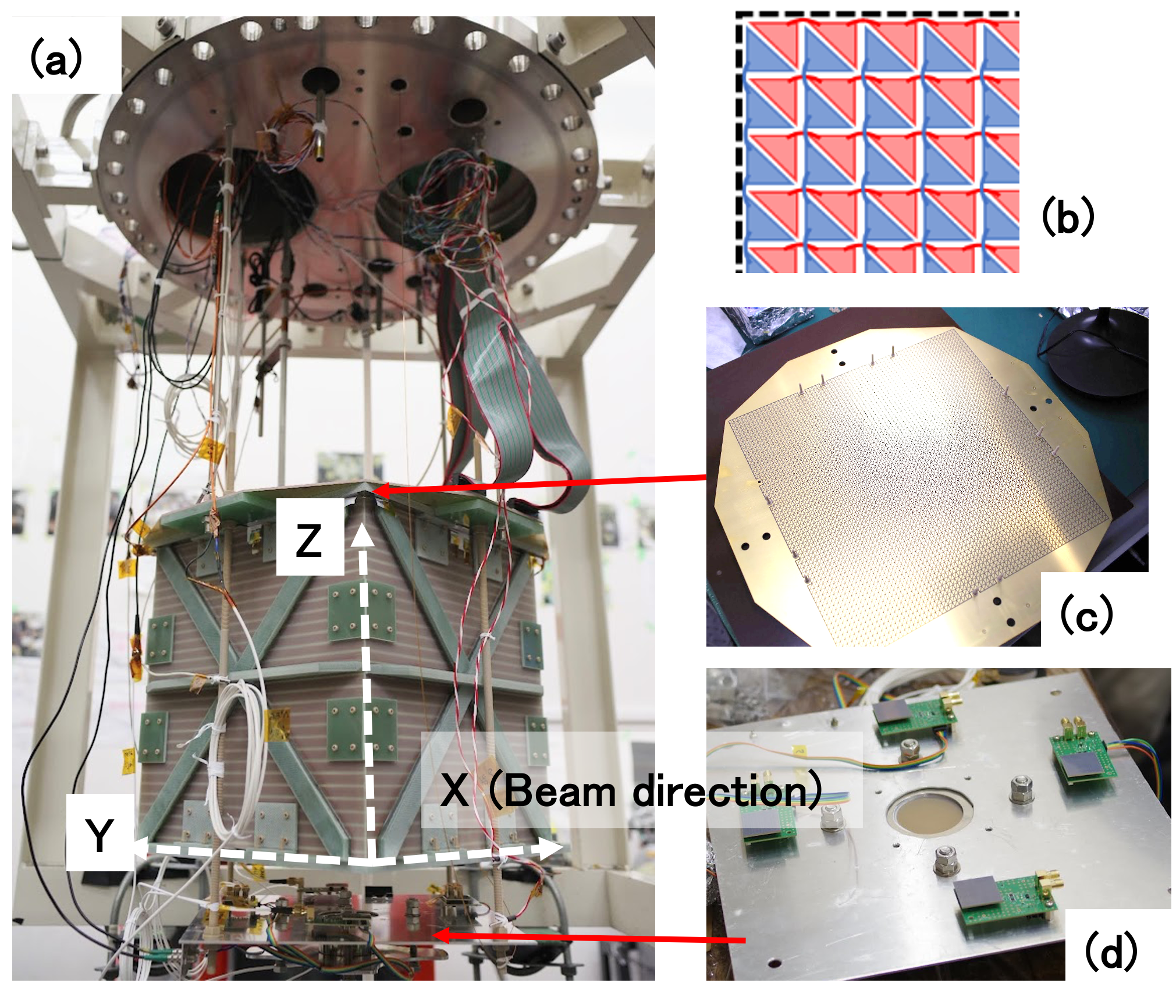}
  \caption{Overview of the LArTPC detector assembly and readout components: (a) Main LArTPC frame suspended beneath the cryostat top flange, highlighting the detector coordinate system ($X$: beam direction, $Y$: transverse horizontal direction, $Z$: vertical drift direction); (b) Schematic pattern of the 2D interleaved $X\text{--}Y$ anode pad structure; (c) Photograph of the assembled 2D anode pad readout plane; (d) Optical sensors (SiPMs/PMTs) and front-end electronics installed at the detector base.}
  \label{fig:lartpc_detector}
\end{figure}

The target detector system deployed at the K1.8BR beamline consists of a single-phase LArTPC housed inside a $200~\text{L}$ stainless-steel vacuum-insulated cryostat. As shown in Fig.~\ref{fig:lartpc_detector}(a), the TPC active volume measures $30 \times 30 \times 30~\text{cm}^3$, corresponding to an active liquid argon mass of approximately $40~\text{kg}$. During operation, a uniform drift electric field of $250~\text{V/cm}$ was applied across the $30~\text{cm}$ drift distance, resulting in a maximum electron drift time of approximately $300~\mu\text{s}$.

To achieve high-spatial-resolution 2D projection imaging and 3D vertex reconstruction of annihilation topologies, a fine-grained two-dimensional (2D) anode readout structure with a $5~\text{mm}$ pitch is mounted at the top of the TPC (Fig.~\ref{fig:lartpc_detector}(b) and (c)). A $100~\mu\text{m}$-thick stainless-steel 2D grid with a $5~\text{mm}$ pitch is placed $1~\text{cm}$ below the anode plane. An induction electric field of $1.5~\text{kV/cm}$ is established in the region between the grid and the anode plane to guide and accelerate drifting electrons onto the anode pads. Ionization electrons passing through the grid induce signals on the $5~\text{mm} \times 5~\text{mm}$ pad array, which is alternately connected along the $X$-axis (aligned with the beam axis) and the $Y$-axis (the transverse horizontal direction), as defined in Fig.~\ref{fig:lartpc_detector}(a), to provide 2D ($X\text{--}Y$) spatial coordinates across 128 readout channels. The charge signals are processed by custom front-end readout electronics boards, named TIGAr (TPC electronics for Ionized signal in Gas and liquid Argon). Each TIGAr board incorporates low-noise LTARS2018K06A ASICs \cite{Tanaka:2020_LTARS} and a Xilinx Artix-7 FPGA, providing preamplification, shaping, and 12-bit ADC digitization at a sampling rate of $2.5~\text{Msps}$, with a maximum DAQ rate of $\sim 100~\text{Hz}$.

In addition to ionization charge readout, the system is equipped with photosensors (PMTs and SiPM arrays, Fig.~\ref{fig:lartpc_detector}(d)) to detect prompt liquid argon scintillation light ($\lambda = 128~\text{nm}$). The inner surfaces of the TPC components are coated with tetraphenyl butadiene (TPB) to shift the vacuum ultraviolet (VUV) photons into the visible regime ($\sim 420~\text{nm}$) \cite{Aoyama:2021xzx}. The fast scintillation signal provides the precise interaction timestamp ($t_0$) required for calculating the $Z$-coordinate (vertical drift direction) from electron drift times, and serves as the primary coincidence trigger in combination with the upstream ToF counters.

\subsection{Beam Exposure and Data Taking in February 2025}
\label{sec:data_taking}
The J-PARC T98 beam test was successfully carried out in February 2025. Following initial detector installation into the preparation area on December 3, 2024, and pre-commissioning runs with LAr filling tests in December and January, the entire LArTPC setup was moved into the K1.8BR experimental area on February 16, 2025. Final pre-cooling and full LAr filling were completed on February 18, 2025. Data collection was conducted during two distinct beam allocations between February 19 and February 21, 2025. The first beam period ran from 22:00 on February 19 to 10:00 on February 20, followed by a second beam period from 12:00 to 24:00 on February 21. 

\begin{figure}[htbp]
\centering
\includegraphics[width=0.95\linewidth]{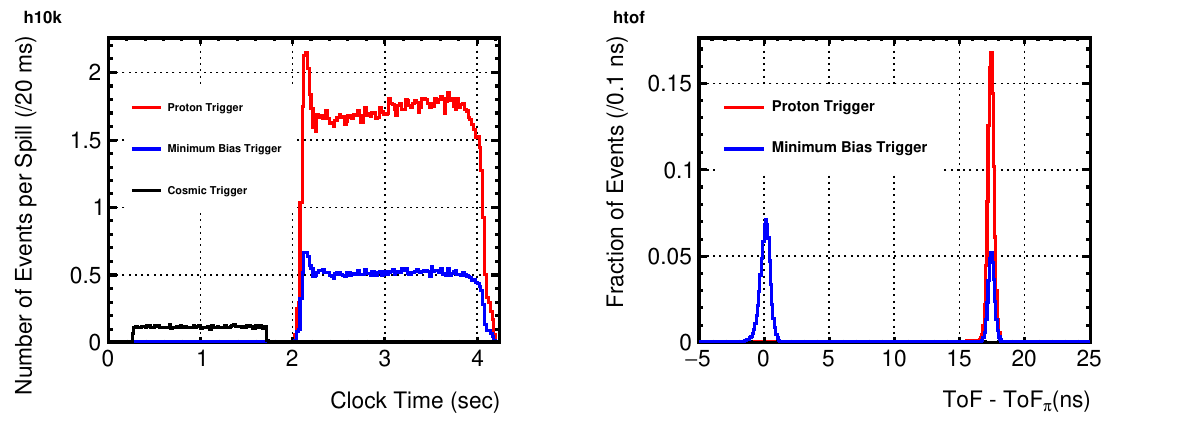}
\caption{Spill-synchronized trigger scheme and beam particle identification: (Left) Time distribution of trigger events within the $4.24~\text{s}$ accelerator spill cycle, showing off-beam cosmic-ray muon triggers (black), online AC-vetoed antiproton triggers labeled as "Proton Trigger" in the plots (red), and prescaled minimum-bias beam triggers (blue); (Right) Relative Time-of-Flight ($\text{ToF} - \text{ToF}_{\pi}$) spectrum comparing the minimum-bias sample (blue) and the AC-vetoed sample (red), demonstrating total suppression of beam pions at $0~\text{ns}$ and pure isolation of antiprotons at $\sim 17.5~\text{ns}$.}
\label{fig:spill_and_tof}
\end{figure}

To efficiently record antiproton interactions while suppressing ambient backgrounds, data acquisition is synchronized with the $4.24~\text{s}$ accelerator spill cycle, as depicted in Fig.~\ref{fig:spill_and_tof} (left). Off-beam timing intervals are utilized to record cosmic-ray muon triggers for detector calibration, whereas on-beam timing is gated with the beam spill. Online particle identification is performed using an Aerogel Cherenkov (AC) counter to veto relativistic pions. Offline selection utilizes the relative Time-of-Flight ($\text{ToF} - \text{ToF}_{\pi}$) measured between the beamline counters, shown in Fig.~\ref{fig:spill_and_tof} (right). While the prescaled minimum-bias beam sample exhibits a dominant pion peak at $0~\text{ns}$, the AC-vetoed trigger configuration suppresses the pion background entirely, yielding a clean and well-isolated antiproton signal centered at $\sim 17.5~\text{ns}$.

During these data-taking periods, the detector was exposed to various beam particle species across different momentum settings to calibrate detector response and evaluate antiproton reaction topologies. Table~\ref{tab:data_summary} summarizes the particle species, beam momentum settings, and total collected statistics. In particular, a high-statistics dataset of $9.8 \times 10^5$ events was successfully accumulated for antiprotons ($\bar{p}$) at the target momentum of $-0.7~\text{GeV}/c$, alongside additional samples at $-1.0~\text{GeV}/c$. Positive proton ($p$), deuteron ($d$), and kaon ($K^-$) samples were also recorded as reference control data to calibrate energy loss ($dE/dx$) and track reconstruction performance.

\begin{table}[htbp]
\centering
\caption{Summary of collected dataset during the J-PARC T98 Phase-2 beam test in February 2025.}
\label{tab:data_summary}
\begin{tabular}{lcc}
\hline\hline
Particle Species & Momentum ($\text{GeV}/c$) & Total Recorded Events \\
\hline
Antiproton ($\bar{p}$) & $-0.7$ & $9.8 \times 10^5$ \\
Antiproton ($\bar{p}$) & $-1.0$ & $6.6 \times 10^5$ \\
Proton ($p$)           & $+0.7$ & $1.6 \times 10^5$ \\
Proton ($p$)/Deuteron ($d$)    & $+1.0$ & $1.5 \times 10^5$ \\
Kaon ($K^-$)           & $-1.0$ & $1.6 \times 10^5$ \\
\hline\hline
\end{tabular}
\end{table}

\section{Event Reconstruction and Data Selection}
\label{sec:reconstruction} 

\subsection{Waveform Processing and 2D Spatial Coordinate Mapping}
\label{sec:waveform_mapping}
The TIGAr readout electronics record the ionization charge signals as time-series ADC waveforms across 128 readout channels ($64$ channels for $X$ and $64$ channels for $Y$). For every trigger event ($t=0$), raw waveforms are acquired over a time window of $[-50~\mu\text{s}, 350~\mu\text{s}]$, fully encompassing the maximum electron drift time ($\sim 300~\mu\text{s}$).

To reconstruct particle trajectories from raw time-series data, the channel waveforms are mapped onto two-dimensional spatial coordinates ($X\text{--}Z$ and $Y\text{--}Z$ planes). Although each readout plane consists of 64 channels, the two outermost channels on both ends (channels 0--1 and 62--63) are physically located outside the TPC field cage boundary and do not collect ionization charges. These outermost channels were intentionally instrumented as reference channels to monitor common-mode noise. After utilizing their baseline fluctuations to estimate and subtract coherent common-mode noise from the active channels, these four channels per plane are excluded from spatial reconstruction.

Consequently, the active physical readout area utilizes 60 channels per axis ($\text{ch}_X, \text{ch}_Y \in [2, 61]$), providing a $300~\text{mm}$ active coverage with a $5~\text{mm}$ pad pitch. The transverse spatial coordinates $X$ and $Y$ are calculated as:
\begin{equation}
    X = (\text{ch}_X - 2) \times 5~\text{mm}, \quad Y = (\text{ch}_Y - 2) \times 5~\text{mm}.
\end{equation}

\begin{figure}[htbp]
\centering
\includegraphics[width=0.95\linewidth]{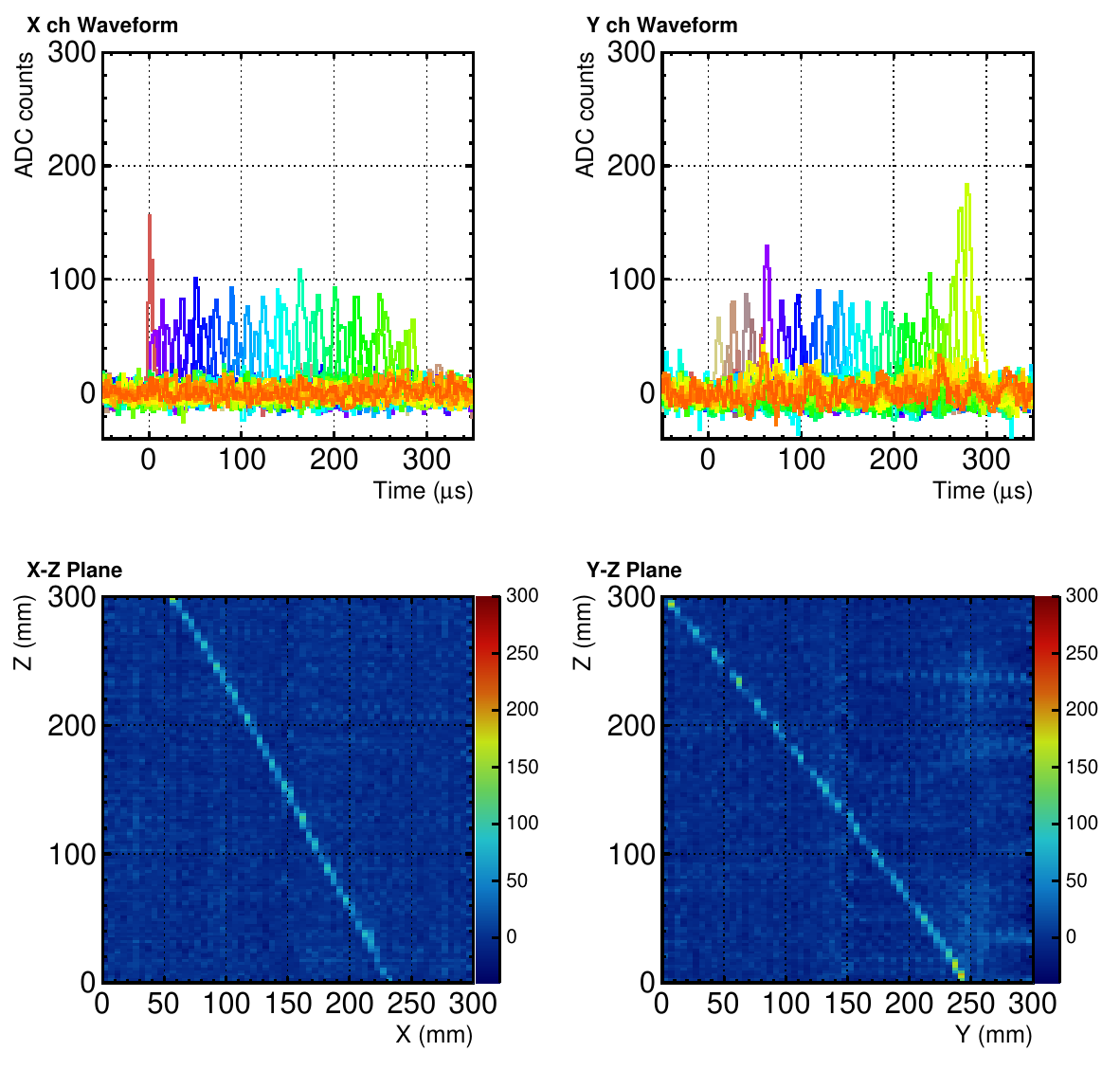}
\caption{Process of converting raw ADC time waveforms into 2D spatial track projections: (Top left) Raw $X$-channel and (Top right) $Y$-channel ADC waveforms recorded in the time window relative to the trigger ($t=0$); (Bottom left) Reconstructed $X\text{--}Z$ plane and (Bottom right) $Y\text{--}Z$ plane 2D spatial energy distributions mapped using $v_{\text{drift}} = 1.05~\text{mm}/\mu\text{s}$, $Z_{\text{anode}} = 300~\text{mm}$, and a $5~\text{mm}$ pad pitch.}
\label{fig:waveform_to_2d}
\end{figure}

Meanwhile, the longitudinal drift coordinate $Z$ is calculated from the electron arrival time $t$. The time origin $t = 0~\mu\text{s}$ corresponds to the arrival of prompt signals at the anode plane situated at $Z = 300~\text{mm}$. Using the drift velocity $v_{\text{drift}} = 1.05~\text{mm}/\mu\text{s}$ under the electric field $E = 250~\text{V/cm}$, the time $t$ is converted into the $Z$-position via:
\begin{equation}
    Z = 300~\text{mm} - v_{\text{drift}} \times t.
\end{equation}
Figure~\ref{fig:waveform_to_2d} illustrates this transformation process for a cosmic-ray muon traversing the LArTPC active volume. The top panels show raw ADC waveforms recorded as a function of time across individual $X$ and $Y$ channels. By combining the $5~\text{mm}$ channel-to-position mapping with the time-to-distance drift conversion, these temporal pulse distributions are reconstructed into clear linear track projections in the $X\text{--}Z$ and $Y\text{--}Z$ planes (bottom panels). The color intensity in each projection plane directly represents the raw ADC count of the corresponding individual readout channel ($X$ or $Y$), without combining signals across orthogonal views.

\begin{figure}[htbp]
\centering
\begin{subfigure}[b]{0.48\textwidth}
    \centering
    \includegraphics[width=\textwidth]{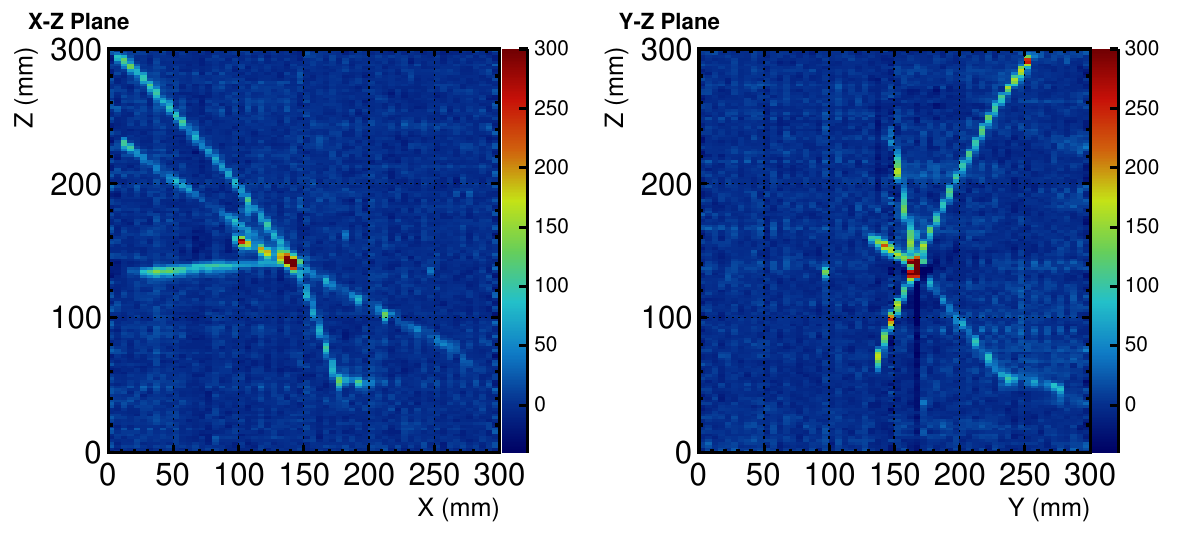}
    \caption{$0.7~\text{GeV}/c~\bar{p}$ (Hadronic star with $\pi/p$ tracks)}
    \label{fig:disp_pbar_hadronic}
\end{subfigure}
\hfill
\begin{subfigure}[b]{0.48\textwidth}
    \centering
    \includegraphics[width=\textwidth]{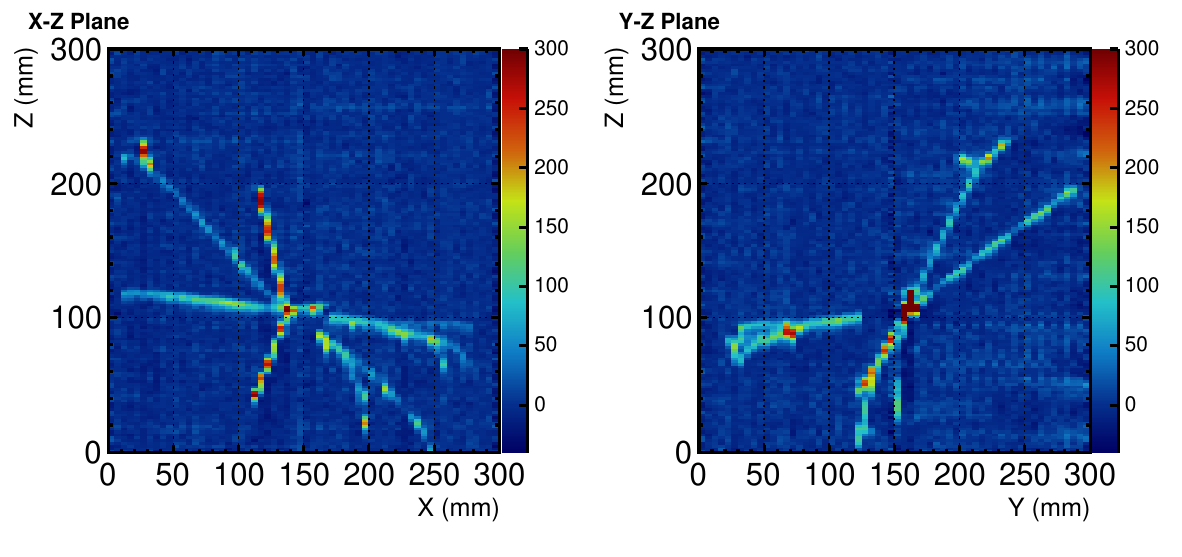}
    \caption{$0.7~\text{GeV}/c~\bar{p}$ (With $\pi^0 \to 2\gamma \to e^+e^-$ shower)}
    \label{fig:disp_pbar_em}
\end{subfigure}

\vspace{0.2cm}

\begin{subfigure}[b]{0.48\textwidth}
    \centering
    \includegraphics[width=\textwidth]{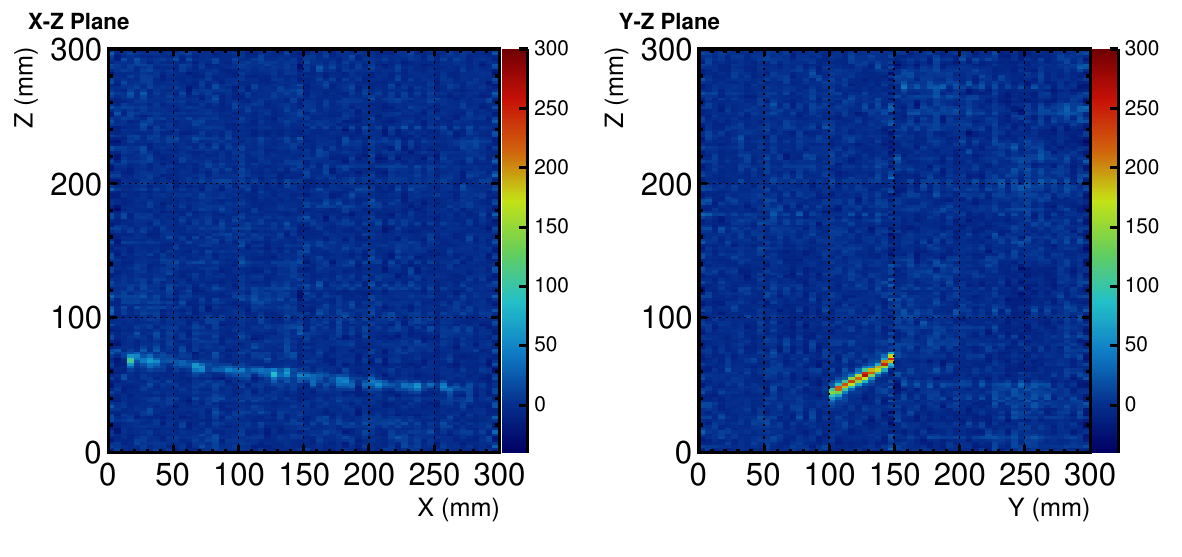}
    \caption{$0.7~\text{GeV}/c~\pi^-$ (Penetrating)}
    \label{fig:disp_pion700}
\end{subfigure}
\hfill
\begin{subfigure}[b]{0.48\textwidth}
    \centering
    \includegraphics[width=\textwidth]{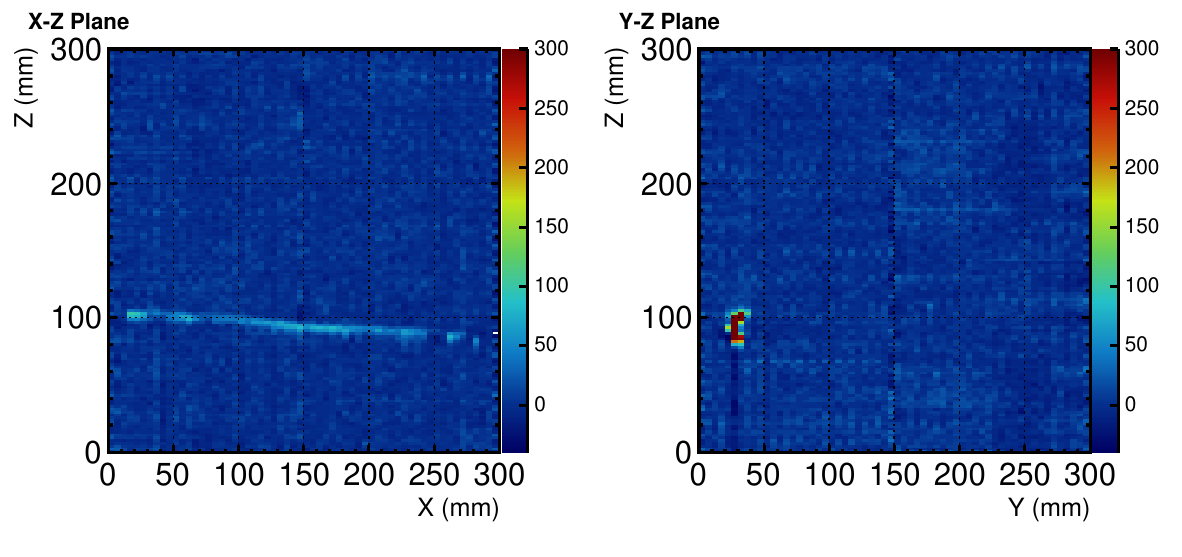}
    \caption{$1.0~\text{GeV}/c~p$ (Penetrating)}
    \label{fig:disp_proton1000}
\end{subfigure}

\vspace{0.2cm}

\begin{subfigure}[b]{0.48\textwidth}
    \centering
    \includegraphics[width=\textwidth]{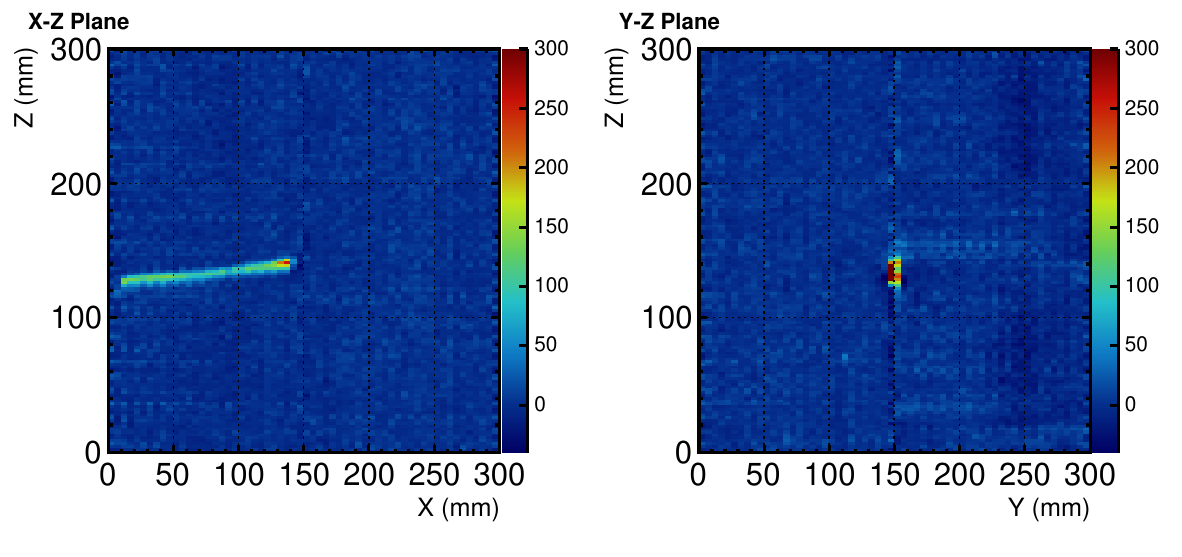}
    \caption{$0.7~\text{GeV}/c~p$ (Stopping)}
    \label{fig:disp_proton700}
\end{subfigure}
\hfill
\begin{subfigure}[b]{0.48\textwidth}
    \centering
    \includegraphics[width=\textwidth]{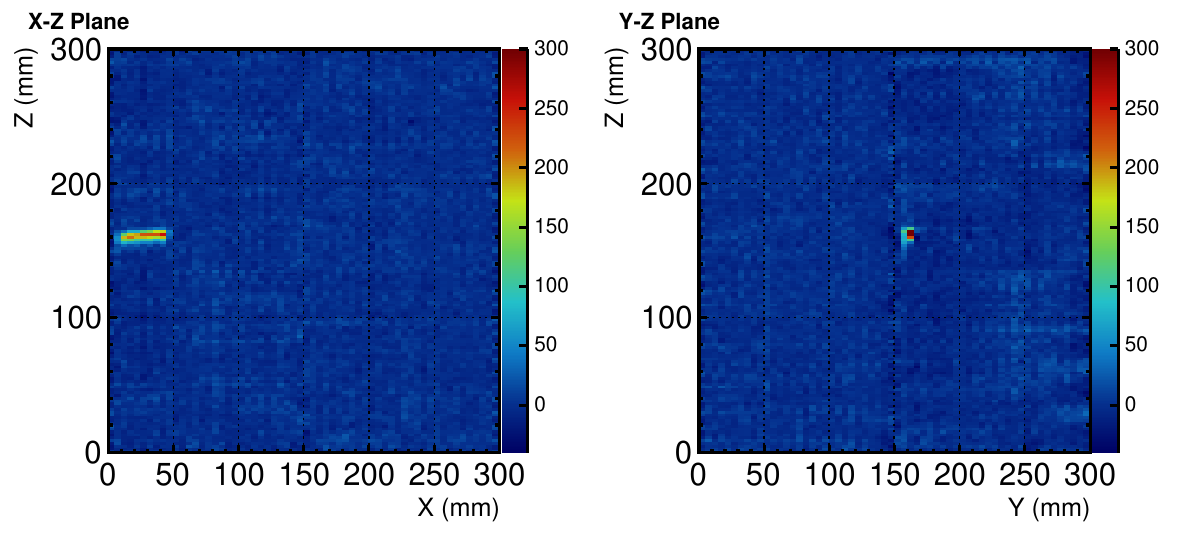}
    \caption{$1.0~\text{GeV}/c~d$ (Stopping)}
    \label{fig:disp_deuteron1000}
\end{subfigure}

\caption{Representative $X\text{--}Z$ and $Y\text{--}Z$ projection event displays of raw ionization signals recorded in the LArTPC: (top left) $0.7~\text{GeV}/c$ antiproton producing hadronic secondaries ($\pi/p$), (top right) $0.7~\text{GeV}/c$ antiproton accompanied by neutral pion decay and electromagnetic conversion ($\pi^0 \to 2\gamma \to e^+e^-$), (middle left) penetrating $0.7~\text{GeV}/c$ pion, (middle right) high-momentum penetrating $1.0~\text{GeV}/c$ proton, (bottom left) stopping $0.7~\text{GeV}/c$ proton showing a Bragg peak, and (bottom right) stopping $1.0~\text{GeV}/c$ deuteron with a high $dE/dx$ short track.}
\label{fig:event_displays}
\end{figure}

Before detailing the automated 3D track reconstruction algorithms, representative 2D event displays recorded during the J-PARC T98 experiment are presented to illustrate the distinct ionization signatures of various particle species inside the LArTPC. Figure~\ref{fig:event_displays} displays the reconstructed $X\text{--}Z$ and $Y\text{--}Z$ projection displays for five different beam and cosmic-ray particles.

The top row highlights two characteristic at-rest antiproton ($\bar{p}$) annihilation topology types at $0.7~\text{GeV}/c$. Figure~\ref{fig:disp_pbar_hadronic} (top left) depicts a multi-prong hadronic annihilation star consisting primarily of outgoing charged pions ($\pi^\pm$) and recoil protons ($p$). Figure~\ref{fig:disp_pbar_em} (top right) shows an annihilation vertex accompanied by electromagnetic activity originating from neutral pion decay ($\pi^0 \to 2\gamma$) followed by pair production ($\gamma \to e^+e^-$), demonstrating the TPC's capability to identify both hadronic tracks and electromagnetic shower components.

The middle row illustrates penetrating particles traversing the active volume with low energy deposition ($dE/dx$). The $0.7~\text{GeV}/c$ pion (middle left) and high-momentum $1.0~\text{GeV}/c$ proton (middle right) enter from $X \sim 0$ and pass through the detector without stopping, providing reference signatures for minimum ionizing or highly relativistic tracks.

The bottom row displays stopping particles that fully deposit their kinetic energy inside the active volume. The $0.7~\text{GeV}/c$ proton (bottom left) penetrates approximately halfway ($\sim 14~\text{cm}$) before coming to rest, exhibiting a clear Bragg peak with enhanced ionization density near the track endpoint. In contrast, the $1.0~\text{GeV}/c$ deuteron (bottom right), having a larger mass and significantly lower velocity ($\beta$) at the same momentum, deposits energy at a much higher rate, resulting in a very short stopping range ($\sim 4\text{--}5~\text{cm}$) and a dense ionization profile. These distinctive topological and calorimetric features demonstrate the clear particle identification (PID) performance of the LArTPC based on track range, topology, and $dE/dx$ profiles.

\subsection{3D Track and Vertex Reconstruction Algorithms}
\label{sec:reco_alg}
\begin{figure}[htbp]
\centering
\includegraphics[width=0.85\linewidth]{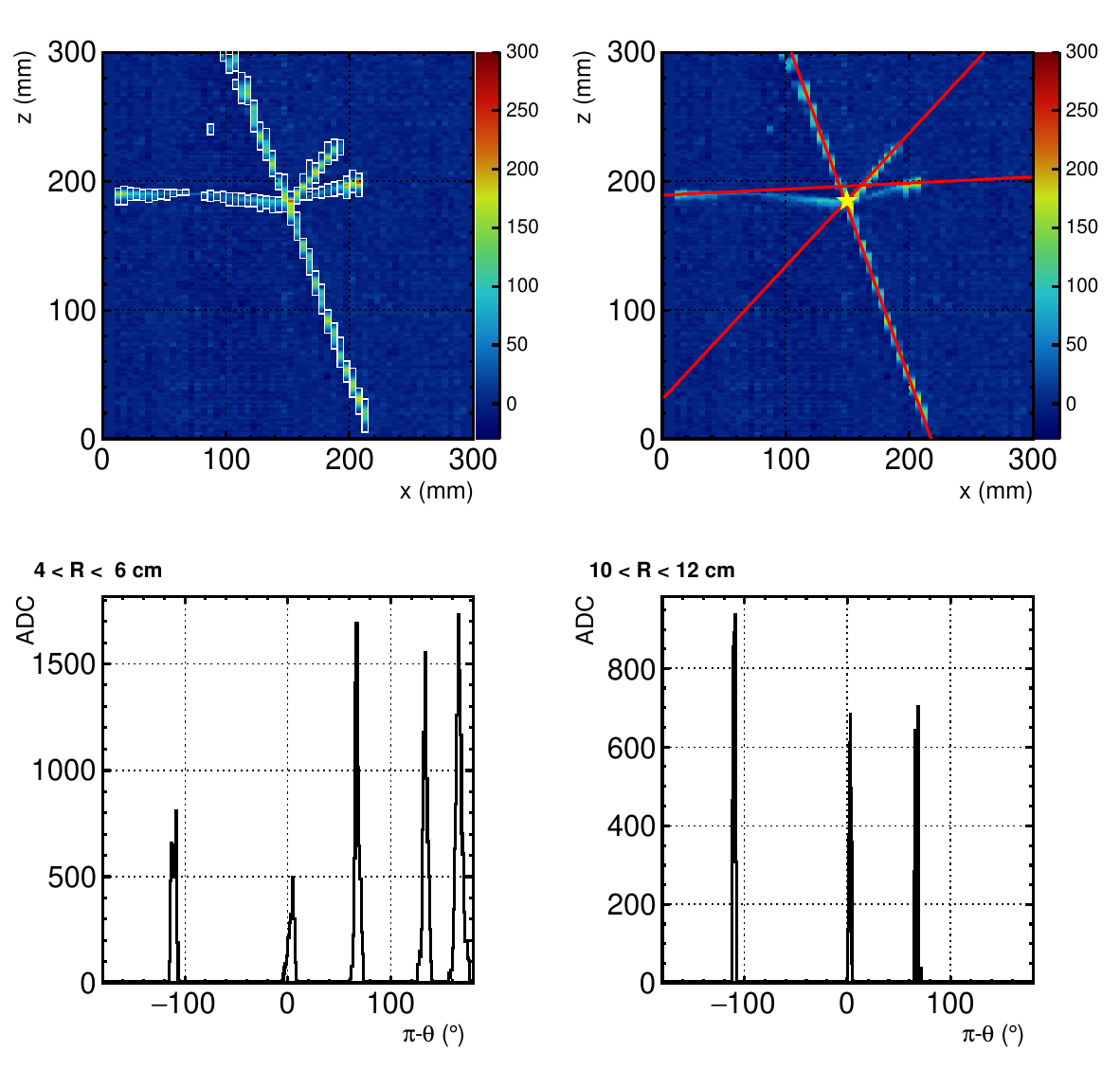} 
\caption{Illustration of the event reconstruction pipeline for an antiproton annihilation event (Event 21): (top left) 2D charge clustering, (top right) linear track identification via Hough transform and vertex finding (yellow star), and (bottom row) angular hit distributions measured in radial slices of $4 < R < 6~\text{cm}$ (bottom left, 5 hits) and $10 < R < 12~\text{cm}$ (bottom right, 3 hits) relative to the vertex.}
\label{fig:reco_steps}
\end{figure}

To systematically analyze the high-statistics antiproton dataset and extract topological features of annihilation "stars", an automated event reconstruction framework was developed. The reconstruction algorithm proceeds through three main steps, as illustrated in Fig.~\ref{fig:reco_steps}:

\begin{enumerate}
    \item \textbf{Clustering}: First, raw ADC waveforms from the 128 readout channels are pedestal-subtracted and integrated to determine the deposited charge. Continuous hit signals on adjacent pads and drift time bins are grouped into 2D clusters in both $X\text{--}Z$ and $Y\text{--}Z$ projection planes. The charge distribution around seed hits is aggregated to define localized energy clusters (Fig.~\ref{fig:reco_steps}, top left).

    \item \textbf{Vertex Finding}: To precisely determine the annihilation endpoint ($\vec{x}_{\text{vtx}}$) of the incoming antiproton, a global chi-square minimization is applied using the reconstructed linear tracks from the Hough transform \cite{Hough:1962,Duda:1972}. For both $X\text{--}Z$ and $Y\text{--}Z$ projection planes, the algorithm minimizes the weighted sum of perpendicular distances from candidate vertex coordinates to each Hough track line. The distance terms are weighted by the corresponding Hough space peak intensities, giving higher significance to prominent particle tracks while sharing a common $Z$-coordinate across both projection views (Fig.~\ref{fig:reco_steps}, top right).

    \item \textbf{Cluster Finding in Radial Slices}: Once the vertex position is established, the active volume surrounding the vertex is divided into concentric radial shell slices ($R$-slices) of width $\Delta R = 2~\text{cm}$. Within each $R$-slice, the energy deposition is projected as a function of the polar angle ($\pi - \theta$) relative to the vertex. Prominent peaks in this angular distribution are reconstructed as \textit{clusters}. This cluster-based approach accommodates both track-like signatures from penetrating charged particles (such as $\pi^\pm$ and protons) and complex energy deposits—most notably electromagnetic showers from $\pi^0 \to \gamma\gamma$ decays that do not point directly to the interaction vertex. As shown in the bottom panels of Fig.~\ref{fig:reco_steps}, an inner slice ($4 < R < 6~\text{cm}$, bottom left) detects five distinct clusters, whereas an outer slice ($10 < R < 12~\text{cm}$, bottom right) shows three clusters as shorter-range particles stop prior to reaching larger radii.
    
\end{enumerate}

\subsection{Energy Scale Calibration using Cosmic-Ray Muons}
\label{sec:muon_calib}
To perform absolute calorimetric measurements and convert raw ADC charge counts into physical deposited energy ($dE/dx$), off-beam cosmic-ray muon events collected during the spill intervals were utilized for detector response calibration. 

\begin{figure}[htbp]
\centering
\includegraphics[width=0.95\linewidth]{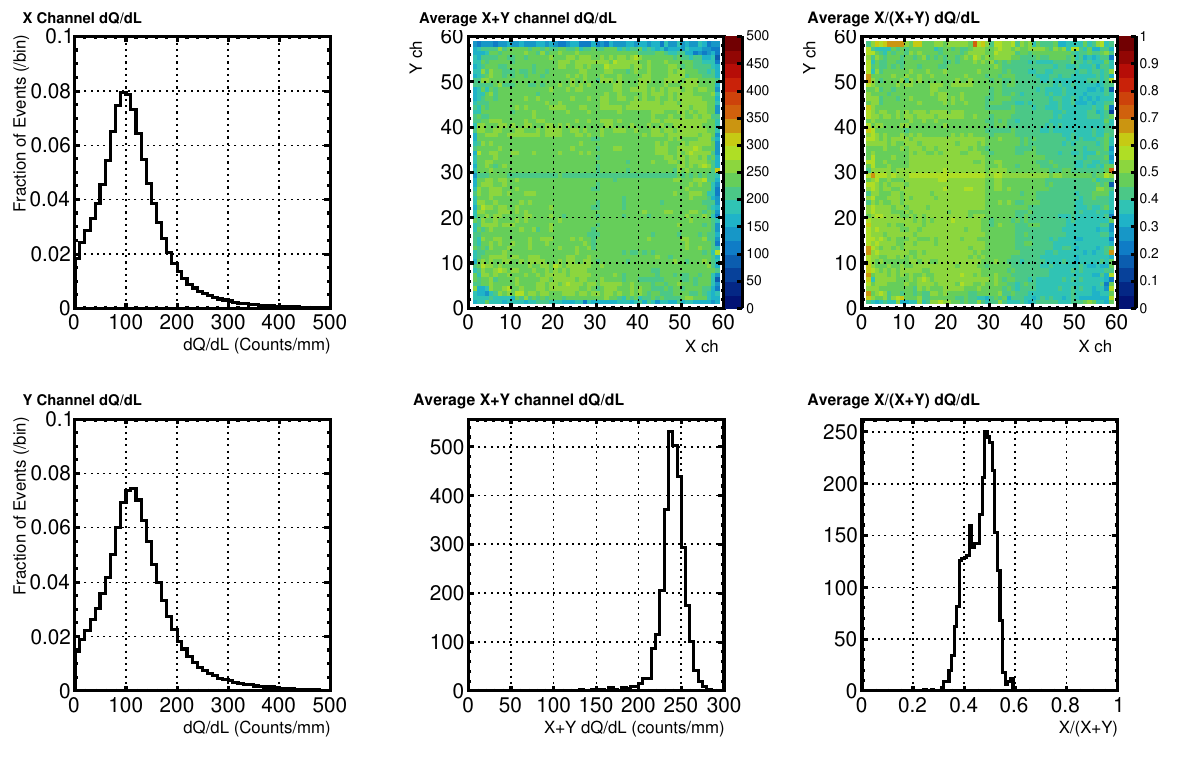}
\caption{Calorimetric calibration and response uniformity evaluated with cosmic-ray muons: (Left column) Measured $dQ/dL$ distributions for $X$ channels (top) and $Y$ channels (bottom); (Middle column) Spatial 2D map of the sum $dQ/dL_{X+Y}$ on $5\times5~\text{mm}^2$ pixel grid (top) and its 1D projection distribution (bottom), where the mean value is calibrated to the muon MIP loss rate ($0.212~\text{MeV/mm}$); (Right column) 2D spatial map (top) and 1D distribution (bottom) of the charge sharing ratio $X/(X+Y)$, showing a subtle spatial modulation due to anode-grid alignment offsets.}
\label{fig:muon_calib}
\end{figure}

Straight single-track events were selected using the 2D clustering and Hough transform algorithms described in Section~\ref{sec:reco_alg}. For each reconstructed track segment traversing a TPC channel, the measured charge $dQ$ is divided by the 3D path length $dL$ to obtain the linear charge density $dQ/dL$ ($\text{ADC counts/mm}$). The resulting 1D $dQ/dL$ spectra for individual $X$ and $Y$ readout channels are shown in the left panels of Fig.~\ref{fig:muon_calib} (top left for $X$, bottom left for $Y$).

To account for channel-to-channel variations and ensure spatial uniformity, the charge contributions from corresponding $X$ and $Y$ channels are combined at each $5\times5~\text{mm}^2$ pixel location. The middle panels of Fig.~\ref{fig:muon_calib} display the 2D spatial map of the combined charge density $dQ/dL_{X+Y}$ across the active anode readout area (top middle) and its overall 1D projection distribution (bottom middle). The average value of the $dQ/dL_{X+Y}$ distribution corresponds to the minimum ionizing particle (MIP) energy deposition rate of cosmic muons in liquid argon, $dE/dx = 0.212~\text{MeV/mm}$. By anchoring this mean $dQ/dL$ response to the theoretical muon MIP scale, an absolute calibration factor converting raw ADC counts to deposited energy ($\text{MeV}$) is established. It should be noted that this linear conversion factor strictly applies to minimum-ionizing signals. Due to electron-ion recombination (quenching) in liquid argon, track segments with high $dE/dx$—such as low-energy stopping protons or nuclear fragments from annihilation—undergo stronger recombination, resulting in an underestimation of deposited energy if uncorrected. The non-linear response for highly ionizing particles is accounted for via non-linear recombination models (Birks' law) in the Monte Carlo simulation (Section~\ref{sec:det_sim}).

Furthermore, the dependence of $dQ/dL$ on the drift distance ($Z$-direction) was evaluated to assess potential ionization charge attenuation caused by electronegative impurities in the liquid argon. No significant $Z$-dependence was observed within the statistical uncertainty, indicating that electron attenuation due to impurities is negligible over the maximum $30~\text{cm}$ drift distance ($300~\mu\text{s}$ drift time) in this dataset.

Finally, the right panels of Fig.~\ref{fig:muon_calib} illustrate the relative charge sharing ratio between the two readout planes, $X / (X+Y)$, presented as a spatial 2D map (top right) and a 1D projection distribution (bottom right). While the 1D projection centers around 0.5, a minor spatial variation is observed in the 2D map: $X/(X+Y)$ systematically decreases toward larger $X$-channel coordinates. This local gradient is attributed to a slight mechanical misalignment ($\sim 1~\text{mm}$) between the anode pad array and the grid plane during detector assembly. To accurately reproduce the detector response, this measured 2D charge-sharing map is directly applied to modulate the simulated charge division in the Monte Carlo framework (Section~\ref{sec:simulation}).

\subsection{Selection of Antiproton Annihilation Candidates}
\label{sec:pbar_selection}

To isolate pure at-rest antiproton reaction events occurring within the LArTPC active volume, a series of sequential offline selection cuts were applied to the reconstructed dataset. 
Table~\ref{tab:event_reduction} summarizes the step-by-step event reduction flow for the $0.7~\text{GeV}/c$ antiproton run, detailing the remaining event yields and cumulative selection efficiencies relative to the total recorded triggers.

\begin{table}[htbp]
\centering
\caption{Summary of the event reduction flow for $0.7~\text{GeV}/c$ antiproton annihilation candidates in the J-PARC T98 dataset.}
\label{tab:event_reduction}
\begin{tabular}{lcc}
\hline\hline
Selection Step & Remaining Events & Cumulative Efficiency (\%) \\
\hline
All recorded triggers                 & 977,251 & 100.0 \\
Online antiproton trigger (AC-vetoed) & 723,587 & 74.0 \\
Offline ToF selection                 & 708,477 & 72.5 \\
Scintillation light pileup rejection  & 323,927 & 33.1 \\
Fiducial vertex selection             & 152,880 & 15.6 \\
\hline\hline
\end{tabular}
\end{table}

\begin{figure}[htbp]
\centering
\includegraphics[width=0.85\linewidth]{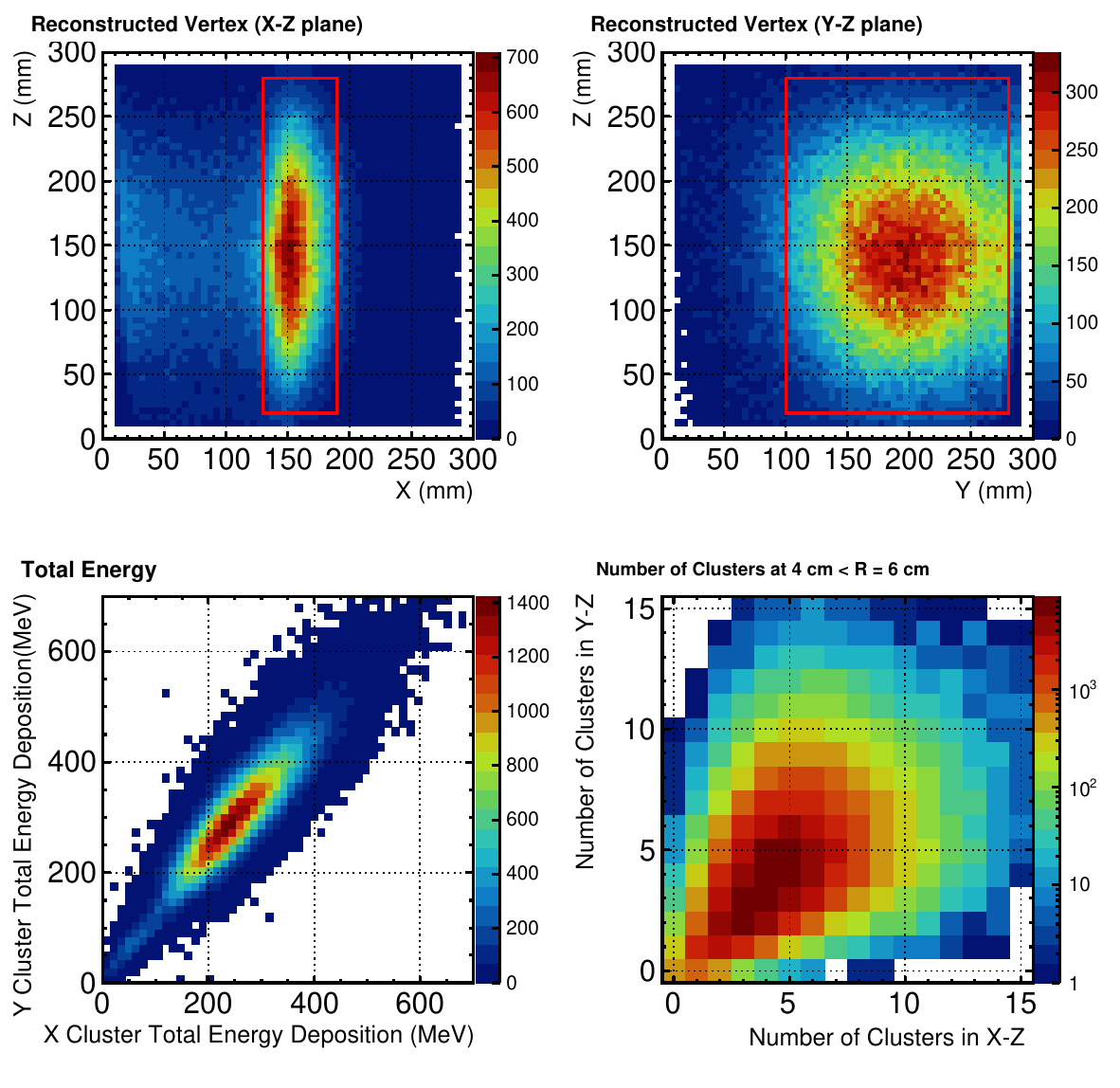}
\caption{Reconstructed antiproton event summary for data: Vertex distributions in the (top left) $X\text{--}Z$ plane and (top right) $Y\text{--}Z$ plane, where the red boxes define the at-rest event selection region; (bottom left) Total cluster energy correlation between $X\text{--}Z$ and $Y\text{--}Z$ projections; (bottom right) Cluster multiplicity correlation at $R = 5~\text{cm}$ from the vertex.}
\label{fig:reco_summary}
\end{figure}

The specific criteria applied at each selection step are as follows:

\begin{enumerate}
    \item \textbf{Online Antiproton Trigger}: From a total of 977,251 recorded trigger events during the $0.7~\text{GeV}/c$ beam run, 723,587 events were selected by requiring the online Aerogel Cherenkov (AC) veto condition to suppress relativistic beam pions.
    
    \item \textbf{Beam Time-of-Flight (ToF) Selection}: An offline ToF window ($\Delta t_{\text{ToF}} \in [16.5, 18.5]~\text{ns}$) centered on the antiproton peak (Fig.~\ref{fig:spill_and_tof}, right) is enforced to reject residual beam pion contaminants and accidental coincidences, leaving 708,477 candidate events.
    
    \item \textbf{Scintillation Light Pileup Rejection Cut}: To eliminate beam pileup artifacts, events are required to have exactly one prompt LAr scintillation light pulse within a time window of $\pm 300~\mu\text{s}$ (corresponding to the full electron drift time). This requirement selects 323,927 clean single-beam interaction events.

    \item \textbf{At-Rest Vertex Selection (Fiducial Cut)}: As shown in the top panels of Fig.~\ref{fig:reco_summary}, $0.7~\text{GeV}/c$ antiprotons lose energy via ionization and come to rest to undergo nuclear capture around $X \sim 150~\text{mm}$. To eliminate boundary edge effects and in-flight reactions, reconstructed interaction vertices are restricted to the fiducial stopping volume defined by $130 < X < 190~\text{mm}$, $100 < Y < 280~\text{mm}$, and $20 < Z < 280~\text{mm}$ (indicated by the red boxes). This final cut yields a pure sample of 152,880 fully reconstructed at-rest annihilation events. \textbf{Based on Monte Carlo simulations (Section~\ref{sec:det_sim}), this selection achieves an estimated at-rest capture purity of $>90\%$, with the remaining fraction consisting of near-stopping in-flight annihilations that are identically modeled in the simulation.}
    
\end{enumerate}

Figure~\ref{fig:reco_summary} demonstrates the self-consistency of the 3D reconstruction across the two independent projection planes for the selected data sample. The lower panels show strong linear correlations between the $X\text{--}Z$ and $Y\text{--}Z$ projections for both the total cluster energy deposition and the cluster multiplicity measured at $R = 5~\text{cm}$ from the vertex.

\section{Monte Carlo Simulations}
\label{sec:simulation}

\subsection{Hadronic Reaction Models: Fritiof, INCL++, and GiBUU}
To evaluate model uncertainties in hadronic interactions, particularly for low-energy antiproton-nucleus ($\bar{p}\text{--}\text{Ar}$) annihilation, three hadronic interaction generators were evaluated and incorporated into the simulation framework:

\begin{itemize}
  \item \textbf{Fritiof Model (\texttt{FTFP\_BERT})}:
  The standard hadronic physics list in Geant4 version 11.1.3 (\texttt{FTFP\_BERT}) treats in-flight antiproton interactions and at-rest antiproton capture through dedicated processes based on the Fritiof model \cite{Andersson:1987_FTF, Agostinelli:2003_Geant4}:
  \begin{itemize}
    \item \textbf{In-flight interactions}: 
    The \texttt{anti\_protonInelastic} process handles in-flight antiproton-nucleus interactions using the Fritiof string model. The excited nuclear remnant is subsequently de-excited using the standard Geant4 pre-compound and nuclear de-excitation modules.

    \item \textbf{At-rest capture}:
    The \texttt{hFritiofCaptureAtRest} process models antiproton annihilation at rest after stopping. The annihilation with a bound nucleon is described by the Fritiof model, followed by the de-excitation of the residual nucleus.
  \end{itemize}

  \item \textbf{Liège Intranuclear Cascade Model (\texttt{FTFP\_INCLXX})}:
  To incorporate refined modeling of antiproton reaction dynamics, we utilize the \texttt{FTFP\_INCLXX} physics list provided in Geant4 version 11.4.2, where the INCL++ framework \cite{Mancusi:2014_INCL, Zharenov:2023_INCL_pbar} is applied to both in-flight interactions and capture at rest:
  \begin{itemize}
    \item \textbf{In-flight interactions}: 
    The \texttt{anti\_protonInelastic} process incorporates the INCL++ model for beam energies up to $10~\text{GeV}$, tracking detailed intranuclear cascades and nuclear medium effects.

    \item \textbf{At-rest capture}:
    The \texttt{hINCLXXCaptureAtRest} process describes antiproton capture and annihilation at rest, using the INCL++ framework to handle multi-nucleon absorption and cluster emission dynamics.
  \end{itemize}

\item \textbf{Giessen Boltzmann-Uehling-Uhlenbeck Model (GiBUU)}: 
GiBUU is a unified transport model solving the Boltzmann-Uehling-Uhlenbeck kinetic equation, rigorously accounting for mean-field potentials, off-shell transport, and Final State Interactions (FSI) such as pion absorption and secondary nucleon knockout inside the argon nucleus \cite{Buss:2012_GiBUU}. Because GiBUU is not natively integrated into Geant4, a custom lookup implementation was constructed:

\begin{enumerate}
    \item Event libraries for $\bar{p}\text{--}\text{Ar}$ annihilations were pre-generated using GiBUU at $25~\mathrm{MeV}$ kinetic energy intervals ($E_k$). Because raw GiBUU event vectors retain all 40 constituent nucleons ($18$ protons and $22$ neutrons) of the $^{40}\text{Ar}$ target nucleus at the end of the transport calculation, an explicit escape filter was applied to the event libraries. Only unbound particles capable of overcoming the nuclear potential barrier were extracted as true secondary emissions, preventing bound nuclear nucleons from being erroneously passed to Geant4 for tracking.

    \item To ensure consistent interaction probabilities between models, the interaction probability per unit kinetic energy, $\text{Prob}(E_k)$ (normalized per $1~\mathrm{MeV}$), was pre-evaluated by simulating $700~\mathrm{MeV}/c$ $\bar{p}$ beam stopping in liquid argon using the standard Fritiof model.

    \item Default Geant4 hadronic inelastic processes for $\bar{p}$ were disabled, leaving only electromagnetic energy loss and multiple scattering active during transport.

    \item At each tracking step with local kinetic energy $E_k$ and energy loss $dE$, the interaction probability was calculated as $P = \text{Prob}(E_k) \times dE$.

    \item Unlike Fritiof, which explicitly separates in-flight inelastic scattering and at-rest capture, GiBUU treats interactions continuously. However, because antiprotons with zero kinetic energy do not trigger interactions in GiBUU, tracking was forcibly terminated once $E_k \le 25~\mathrm{MeV}$ to prevent non-physical penetration, and annihilation was evaluated using the $25~\mathrm{MeV}$ GiBUU library.

    \item Upon interaction, the primary antiproton was replaced with secondary products sampled from the corresponding filtered GiBUU library, followed by standard Geant4 tracking of these produced secondaries.
\end{enumerate}

\end{itemize}

\begin{figure}[htbp]
\centering
\includegraphics[width=0.85\linewidth]{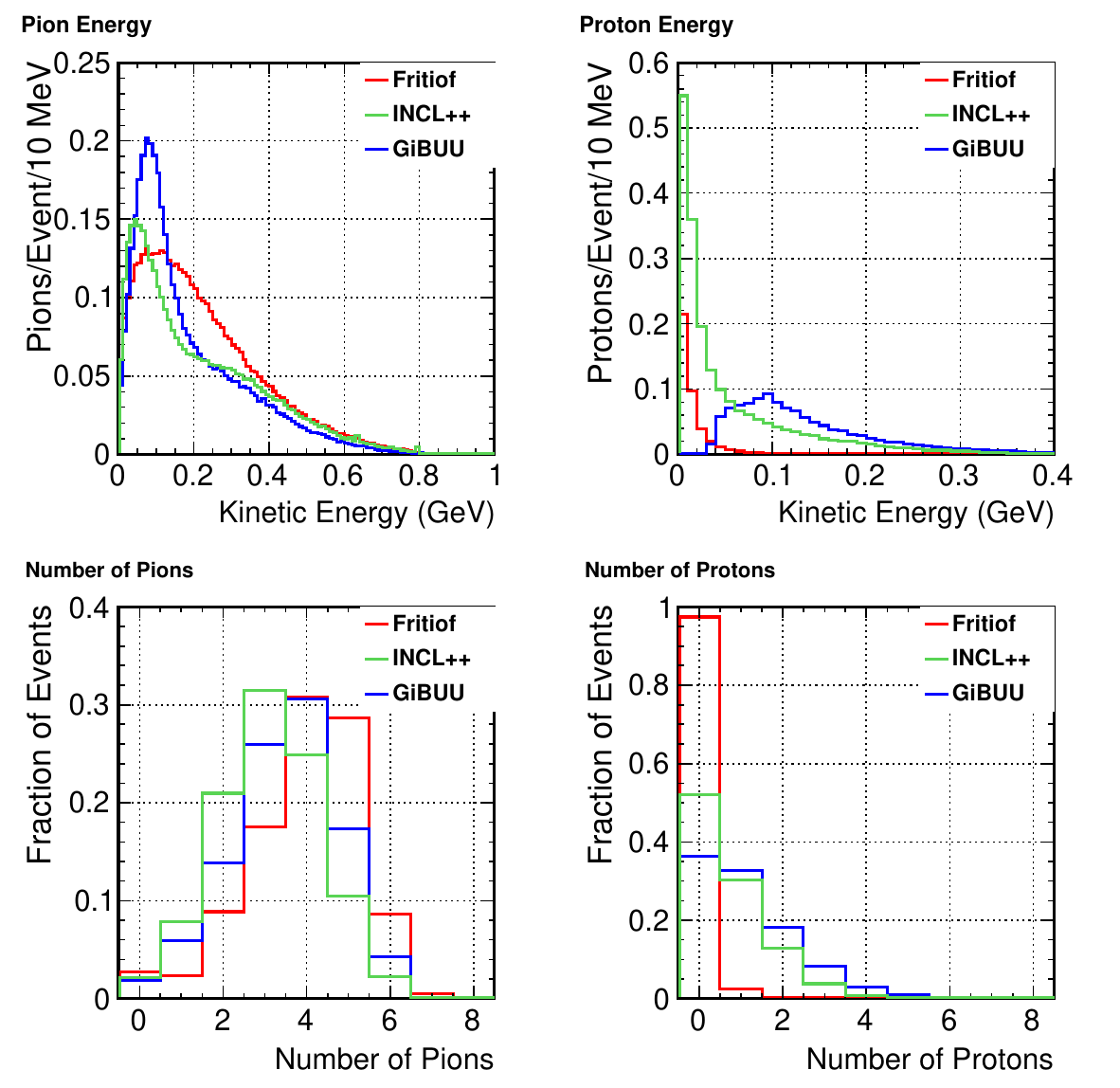}
\caption{Generator-level comparisons of secondary particle spectra from $\bar{p}\text{--}\text{Ar}$ annihilation predicted by Fritiof (red), INCL++ (green), and GiBUU (blue): Kinetic energy distributions per event for total pions ($\pi^+ + \pi^- + \pi^0$, top-left) and protons (top-right); and multiplicity distributions per event for pions (bottom-left) and protons (bottom-right).}
\label{fig:generator_comparison}
\end{figure}

Significant differences in secondary particle production at the generator level are highlighted in Fig.~\ref{fig:generator_comparison}:
\begin{itemize}
    \item \textbf{Proton Emission and Multiplicity}: 
    Significant model discrepancies emerge in both the kinetic energy spectra (top-right) and multiplicity distributions (bottom-right) of secondary protons. The Fritiof model yields negligible proton emission, with $\sim 97\%$ of events producing zero protons and energy loss restricted below $50~\text{MeV}$. Conversely, both cascade/transport models predict substantial proton knockouts resulting from Final State Interactions (FSI). INCL++ exhibits a strong peak at $0~\text{MeV}$ with a soft high-energy tail extending up to $0.4~\text{GeV}$, producing $\ge 1$ proton in $\sim 48\%$ of events. GiBUU predicts the highest overall proton multiplicity ($\ge 1$ proton in $\sim 64\%$ of events, with tails extending up to $4\text{--}5$ protons) and a distinct kinetic energy peak around $E_k \sim 100~\text{MeV}$. The deficit of sub-barrier protons ($E_k \sim 0~\text{MeV}$) in GiBUU is directly caused by the explicit potential barrier escape filter.

    \item \textbf{Pion Production, Energy, and Absorption}: 
    The pion kinetic energy spectra (top-left) and multiplicity distributions (bottom-left) reflect varying degrees of in-medium pion absorption. Fritiof predicts the highest pion multiplicity (peaking at $5$ pions per event) and a broader kinetic energy spectrum extending above $0.2~\text{GeV}$ due to minimal FSI. In contrast, both INCL++ and GiBUU account for pion absorption channels that convert mesonic energy into nucleon knockouts. GiBUU exhibits a prominent low-energy pion peak around $100~\text{MeV}$ with a multiplicity peaking at $4$ pions. INCL++ predicts the lowest overall pion multiplicity (peaking at $3\text{--}4$ pions per event), consistent with its strong pion absorption rate.
\end{itemize}

\subsection{Incident Beam Simulation and Input Profile}
\label{sec:beam_profile}
To accurately model the antiproton propagation and subsequent interaction topologies inside the LArTPC, the Monte Carlo (MC) simulation incorporates realistic incident beam profiles directly measured by the upstream beamline detectors during the experiment.

\begin{figure}[htbp]
\centering
\includegraphics[width=1.0\linewidth]{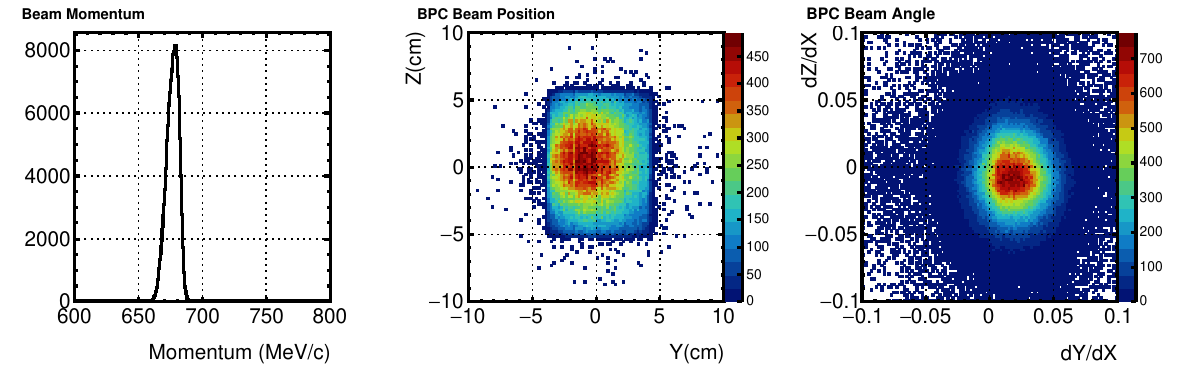}
\caption{Measured beam profiles used as primary particle input for Monte Carlo simulations: (Left) Beam momentum spectrum reconstructed via the D5 dipole magnet curvature; (Middle) Transverse position distribution ($Y\text{--}Z$) and (Right) directional angles ($dY/dX$ vs. $dZ/dX$) measured by the BPC counters located $2.6~\text{m}$ upstream from the LArTPC center.}
\label{fig:beam_profile}
\end{figure}

The kinematic parameters of individual beam particles—specifically momentum, spatial entrance position, and incident direction—were extracted event-by-event using the beamline instrumentation setup. The left panel of Fig.~\ref{fig:beam_profile} shows the measured incident beam momentum spectrum, reconstructed from particle trajectory curvatures in the D5 dipole magnet, exhibiting a sharp momentum distribution centered around $\sim 680~\text{MeV}/c$. 

Furthermore, the spatial position and angular divergence of the beam were reconstructed using a Beam Profile Chamber (BPC) positioned $2.6~\text{m}$ upstream from the LArTPC center. The middle and right panels of Fig.~\ref{fig:beam_profile} display the transverse position distribution ($Y\text{--}Z$ plane) and incident directional angles ($dY/dX$ vs. $dZ/dX$), respectively. In the Monte Carlo generation pipeline, initial primary antiproton candidates are generated via pseudo-random sampling from these empirical momentum, position, and angular distributions. This data-driven beam initialization ensures precise reproduction of the spatial distribution and stopping position of primary beam particles inside the LArTPC active volume.

\subsection{Detector Simulation and Electronics Response Framework}
\label{sec:det_sim}
To directly compare Monte Carlo (MC) predictions with experimental data, a comprehensive detector simulation pipeline was developed. The framework reproduces the full chain of physical processes, from primary particle energy deposition inside the liquid argon to the final digital ADC waveform output.

\begin{enumerate}
    \item \textbf{Geant4 Detector Geometry and Energy Deposition}: 
    The geometry of the T98 LArTPC—including the $200~\text{L}$ cryostat, active liquid argon volume ($30\times30\times30~\text{cm}^3$), 2D grid structure, and anode pad planes—was constructed within the Geant4 simulation framework. Primary beam particles, initialized with the measured momentum and spatial profiles (Section~\ref{sec:beam_profile}), are propagated through the detector geometry, and step-by-step ionization energy depositions ($dE/dx$) are computed with a fixed step length of 0.1~mm.

    \item \textbf{Ionization Charge and Recombination (Birks' Law)}:
    The total initial charge generated along each step is calculated using the average ionization energy for LAr ($W_{\text{ion}} = 23.6~\text{eV}$) \cite{Miyajima:1974_Wion}. Electron-ion recombination effects under the applied drift electric field of $E_{\text{drift}} = 0.25~\text{kV/cm}$ ($250~\text{V/cm}$) are modeled using Birks' law:
    \begin{equation}
        R = \frac{A}{1 + k_B \cdot (dE/dx)/ E_{\text{drift}}},
    \end{equation}
    where $R$ is the recombination factor, and $A = 0.800$ and $k_B = 0.0486~\text{kV/cm}\cdot\text{g}/(\text{cm}^2\cdot\text{MeV})$ are empirical Birks parameters \cite{ICARUS:2004koz}. Here, $dE/dx$ represents the stopping power expressed in $\text{MeV}/(\text{g/cm}^2)$, evaluated for each Geant4 step as the deposited energy ($\text{MeV}$) divided by the product of the step length ($\text{cm}$) and the density of the liquid argon ($\rho = 1.40~\text{g/cm}^3$).

    \item \textbf{Electron Drift, Diffusion, Induction, and Spatial Response Mapping}:
    Surviving ionization electrons are drifted along the uniform electric field lines toward the anode plane at the electron drift velocity ($\sim 1.05~\text{mm}/\mu\text{s}$). Spatial blurring during transport is accounted for by applying transverse and longitudinal electron diffusion with diffusion coefficients of $D_T = 9.5~\text{cm}^2/\text{s}$ and $D_L = 6.3~\text{cm}^2/\text{s}$, respectively \cite{Li:2016_diffusion, Shibamura:1979_diffusion}. Upon passing through the 2D grid plane into the $1.5~\text{kV/cm}$ induction field, the induced current on each readout channel was calculated from the time variation of the solid angle subtended by the corresponding electrode, based on the Shockley–Ramo theorem. To account for position-dependent charge collection non-uniformities without requiring complex 3D track matching on data, the empirical $X/(X+Y)$ charge-sharing map measured from cosmic muons (Section~\ref{sec:muon_calib}) is applied to modulate the charge division between the $X$- and $Y$-readout pads as a function of 3D hit position.
    
    \item \textbf{TIGAr Electronics Response and Digitization}:
    The induced current signals on each readout channel are convolved with the transfer function of the custom TIGAr front-end electronics, incorporating preamplification, ASIC shaping time, and electronic noise. Finally, the shaped analog signals are digitized into 12-bit ADC waveforms at a sampling rate of $2.5~\text{Msps}$, generating raw Monte Carlo datasets in an identical data format to the real experimental runs.
\end{enumerate}

\begin{figure}[htbp]
\centering
\includegraphics[width=0.85\linewidth]{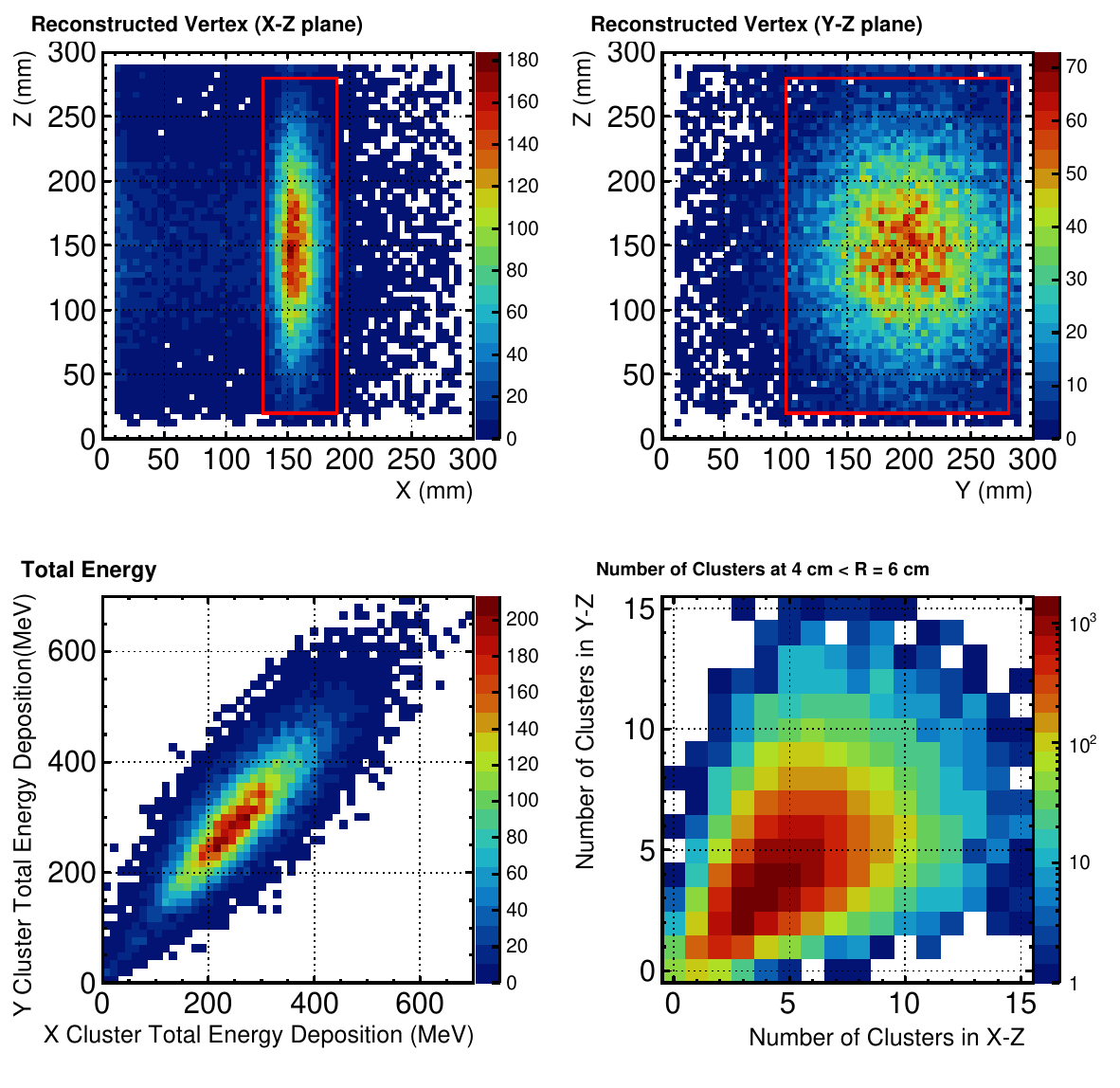}
\caption{Reconstructed antiproton event summary for the GiBUU MC sample: Vertex distributions in the (top left) $X\text{--}Z$ plane and (top right) $Y\text{--}Z$ plane, where the red boxes define the at-rest event selection region; (bottom left) Total cluster energy correlation between $X\text{--}Z$ and $Y\text{--}Z$ projections; (bottom right) Cluster multiplicity correlation at $R = 5~\text{cm}$ from the vertex.}
\label{fig:mc_reco_summary}
\end{figure}

By processing the simulated digital waveforms through the exact same track reconstruction and vertex-finding algorithms used for real data (Section~\ref{sec:reco_alg}), an unbiased Data-MC comparison is achieved.
Figure~\ref{fig:mc_reco_summary} presents the corresponding reconstruction summary for the simulated $\bar{p}$ annihilation events generated with the GiBUU model.
The simulated distributions show good agreement with the measured data (Figure~\ref{fig:reco_summary}), successfully reproducing both the spatial vertex topology and the strong linear correlations between the two orthogonal projection planes in energy deposition and cluster multiplicity.
In contrast, inspecting the vertex distribution for the INCL++ simulation indicates that low-energy in-flight inelastic interactions prior to stopping are scarcely generated, resulting in vertices concentrated almost exclusively in the at-rest capture region. However, because the fiducial vertex selection ($130 < X < 190~\text{mm}$) predominantly isolates at-rest annihilation events ($>90\%$ purity), this deficit in upstream in-flight modeling has a negligible impact on the final selected sample and the overall Data-MC comparisons.

Prior to evaluating the complex $\bar{p}\text{--}\text{Ar}$ annihilation topologies, the baseline calorimetric energy scale and detector response simulation were validated using a control sample of $1.0~\text{GeV}/c$ protons traversing the active LAr volume. Since penetrating protons in this momentum range deposit energy predominantly through electromagnetic processes without undergoing inelastic hadronic interactions, the MC simulation relied solely on standard Geant4 ionization energy loss ($dE/dx$) modeling.

\begin{figure}[htbp]
\centering
\includegraphics[width=0.65\linewidth]{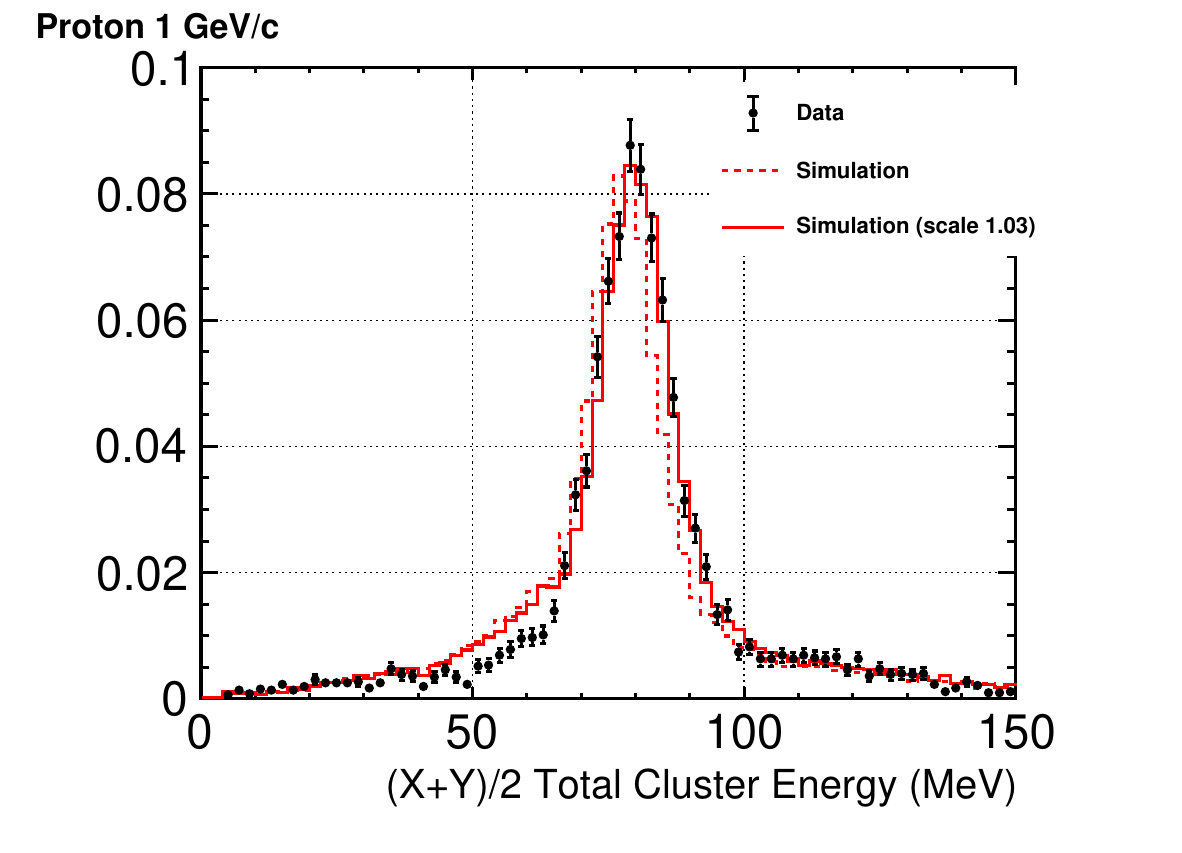}
\caption{Comparison of total reconstructed cluster energy (averaged across the $X$ and $Y$ readout views) for penetrating $1.0~\text{GeV}/c$ control protons between experimental data (black markers) and high-statistics Geant4 simulation. The nominal simulation is shown by the dashed red line, while the solid red line represents the simulation scaled by a factor of 1.03.}
\label{fig:proton10_energy}
\end{figure}

Figure~\ref{fig:proton10_energy} presents the total reconstructed energy deposition, evaluated as the mean energy between the $X$- and $Y$-readout views, for penetrating $1.0~\text{GeV}/c$ beam protons, comparing experimental data with a high-statistics Geant4 MC simulation.
The nominal simulation (dashed red line) reproduces the overall spectral shape and energy loss peak around $\sim 80~\text{MeV}$ reasonably well, but exhibits a minor energy scale discrepancy of approximately $3\%$ relative to data.
Applying a $1.03$ scaling factor to the simulation energy (solid red line) yields excellent agreement with the measured distribution.

This $3\%$ energy scale difference is attributed to subtle detector response effects that are not fully captured by the baseline calibration. Specifically, the absolute energy scale was calibrated using cosmic-ray muons, which are Minimum Ionizing Particles (MIPs) entering the TPC from isotropic directions. In contrast, the control protons propagate strictly along the beam axis and possess higher local $dE/dx$, introducing residual uncertainties in angle-dependent charge collection and recombination effects. Consequently, an overall $3\%$ uncertainty is assigned as a systematic uncertainty on the calorimetric energy scale for the subsequent $\bar{p}$ annihilation observables.

\section{Results and Discussion}
\label{sec:results}

\subsection{Data and Monte Carlo Model Comparisons}
\label{sec:datamc_comparison}

To evaluate the performance of the hadronic interaction models, the selected $\bar{p}\text{--}\text{Ar}$ annihilation candidate events from experimental data are directly compared with full detector simulations employing three generator models: Fritiof, INCL++, and GiBUU. In all figures, experimental data points include combined statistical uncertainties and the $3\%$ calorimetric energy scale systematic uncertainty assigned from control proton validations. To quantitatively evaluate the goodness-of-fit between data and each hadronic generator, reduced chi-square values ($\chi^2/\text{ndf}$) were calculated across all observables, as summarized in Table~\ref{tab:chi2_summary}. For observables combined across multiple radial shells as well as the grand total, the combined reduced chi-square is evaluated as $\sum \chi^2 / \sum \text{ndf}$.

\begin{table}[htbp]
\centering
\caption{Quantitative comparison of reduced chi-square ($\chi^2/\text{ndf}$) values between experimental data and Monte Carlo predictions for total deposited energy, radial energy profiles ($\Sigma R$-slices), spatial cluster multiplicities ($\Sigma R$-slices), and the grand total across all distributions. Combined values represent $\sum \chi^2 / \sum \text{ndf}$.}
\label{tab:chi2_summary}
\begin{tabular}{lcccc}
\hline\hline
Model & Total Energy & Energy Slices ($\Sigma R$) & Multiplicity Slices ($\Sigma R$) & Grand Total \\
\hline
Fritiof  & 1.7 & 3.7 & 7.0 & 4.0 \\
INCL++   & 1.7 & 9.3 & 1.2 & 6.5 \\
GiBUU    & 0.9 & 3.1 & 0.5 & 2.2 \\
\hline\hline
\end{tabular}
\end{table}

\begin{figure}[htbp]
\centering
\includegraphics[width=0.65\linewidth]{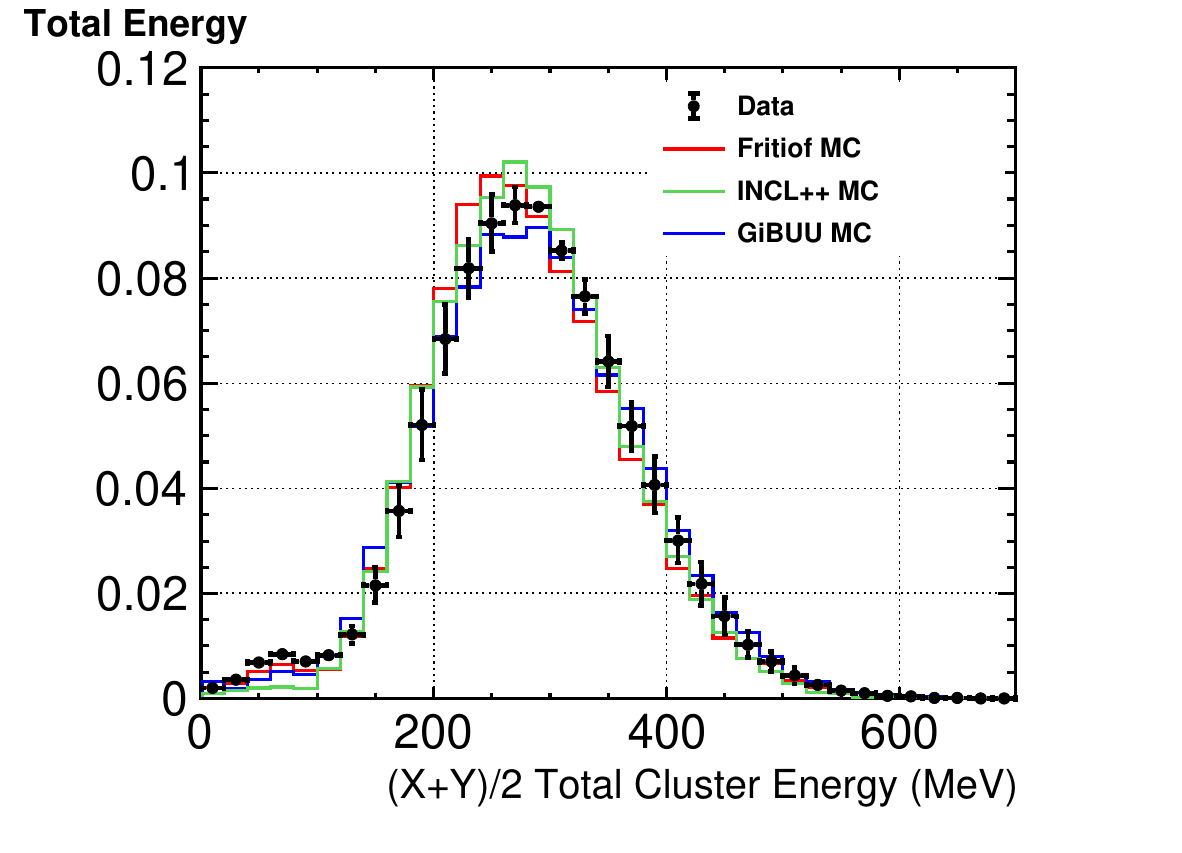}
\caption{Comparison of total deposited energy (averaged across the $X$ and $Y$ readout views) per annihilation event between experimental data (black markers with statistical and $3\%$ systematic uncertainties), Fritiof MC (red), INCL++ MC (green), and GiBUU MC (blue).}
\label{fig:total_energy}
\end{figure}

\textbf{Total Deposited Energy Distribution}: 
Figure~\ref{fig:total_energy} presents the total reconstructed energy deposited within the event window, evaluated as the mean energy between the $X$ and $Y$ readout views. While all three hadronic models broadly reproduce the overall distribution centered around $\sim 280~\text{MeV}$, subtle systematic shifts in the mean deposited energy are observed among the generators, following the hierarchy of $\text{Fritiof} < \text{GiBUU} < \text{INCL++}$. As detailed in Table~\ref{tab:chi2_summary}, GiBUU achieves the best quantitative agreement with $\chi^2/\text{ndf} = 0.9$, whereas Fritiof ($\chi^2/\text{ndf} = 1.7$) and INCL++ ($\chi^2/\text{ndf} = 1.7$) exhibit slight shape discrepancies.

\begin{figure}[htbp]
\centering
\includegraphics[width=0.98\linewidth]{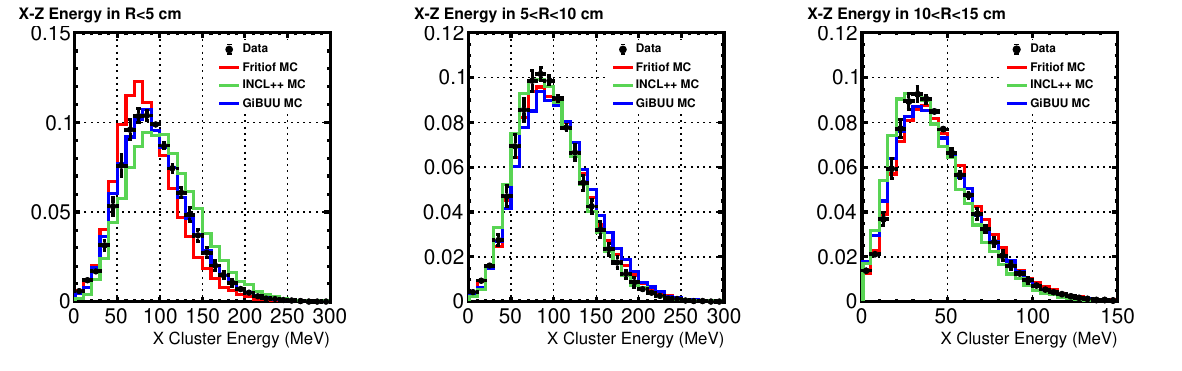}
\caption{Deposited energy spectra reconstructed within different radial distances $R$ from the annihilation vertex: (Left) $R < 5~\text{cm}$, (Middle) $5 < R < 10~\text{cm}$, and (Right) $10 < R < 15~\text{cm}$. Data points include combined statistical and $3\%$ scale systematic uncertainties.}
\label{fig:radial_energy}
\end{figure}

\textbf{Radial Energy Profiles ($R$-slices)}: 
Figure~\ref{fig:radial_energy} shows the reconstructed cluster energy deposited within three concentric radial shells around the annihilation vertex ($R < 5~\text{cm}$, $5 < R < 10~\text{cm}$, and $10 < R < 15~\text{cm}$). The combined goodness-of-fit across these energy slices (Table~\ref{tab:chi2_summary}, second column: $\chi^2/\text{ndf} = 3.1$ for GiBUU, $3.7$ for Fritiof, and $9.3$ for INCL++) highlights critical differences in modeling vertex-adjacent energy loss:
\begin{itemize}
    \item In the innermost region ($R < 5~\text{cm}$), where low-energy knockout protons dominate the visible energy deposition (heavy nuclear fragments contribute negligibly due to severe recombination quenching in liquid argon), clear differences in average energy loss emerge: Fritiof yields the lowest mean energy loss due to a lack of intranuclear cascade knockouts, whereas INCL++ significantly overestimates local energy deposition with an extended high-energy tail (driving its elevated $\chi^2/\text{ndf} = 9.3$). GiBUU lies intermediate between the two, closely tracking the experimental profile.
    \item At intermediate radii ($5 < R < 10~\text{cm}$), all three models demonstrate reasonable agreement with experimental data within uncertainties, with Fritiof and INCL++ exhibiting slightly lower mean energy loss compared to GiBUU.
    \item In the outer region ($10 < R < 15~\text{cm}$), dominated by penetrating charged pions, $e^+e^-$ pairs from $\pi^0 \to 2\gamma$ decay photons, and energetic protons, all models show good overall consistency with data, with INCL++ predicting a subtly lower average energy loss relative to Fritiof and GiBUU.
\end{itemize}

\begin{figure}[htbp]
\centering
\includegraphics[width=0.98\linewidth]{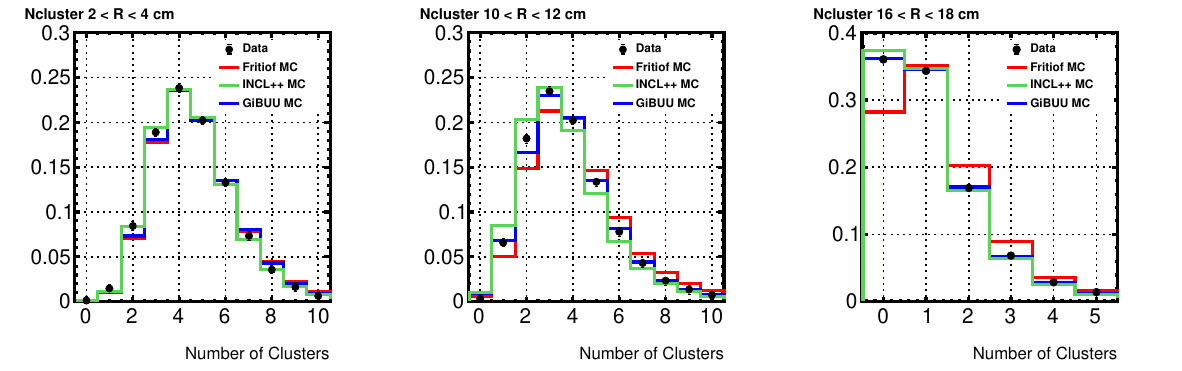}
\caption{Reconstructed hit cluster multiplicity distributions evaluated within specific radial shells from the vertex: (Left) $2 < R < 4~\text{cm}$, (Middle) $10 < R < 12~\text{cm}$, and (Right) $16 < R < 18~\text{cm}$. Data points include statistical and $0.5\%$ energy-threshold systematic uncertainties.}
\label{fig:cluster_multiplicity}
\end{figure}

\textbf{Spatial Cluster Multiplicities ($R$-slices)}: 
Figure~\ref{fig:cluster_multiplicity} displays the distribution of reconstructed hit clusters in representative radial slices ($2 < R < 4~\text{cm}$, $10 < R < 12~\text{cm}$, and $16 < R < 18~\text{cm}$). The combined goodness-of-fit across the cluster multiplicity slices clearly differentiates the models (Table~\ref{tab:chi2_summary}, third column: $\chi^2/\text{ndf} = 0.5$ for GiBUU, $1.2$ for INCL++, and $7.0$ for Fritiof):
\begin{itemize}
    \item Near the core region ($2 < R < 4~\text{cm}$), all three models capture the cluster multiplicity distribution well, with INCL++ and GiBUU matching the peak particularly well.
    \item At larger radial distances ($10 < R < 12~\text{cm}$ and $16 < R < 18~\text{cm}$), GiBUU demonstrates excellent empirical agreement ($\chi^2/\text{ndf} = 0.5$). INCL++ slightly overpredicts the low-multiplicity bins, whereas Fritiof exhibits a severe excess in the high-multiplicity tail due to minimal pion absorption, resulting in an elevated $\chi^2/\text{ndf}$ of $7.0$.
\end{itemize}

In summary, as quantitatively summarized by the grand total reduced chi-square values in Table~\ref{tab:chi2_summary} ($\chi^2/\text{ndf}_{\text{Grand}} = 2.2$ for GiBUU, $4.0$ for Fritiof, and $6.5$ for INCL++), GiBUU provides the most consistent overall description of at-rest $\bar{p}\text{--}\text{Ar}$ interactions, successfully balancing vertex proton knockouts and large-radius pion absorption.

\subsection{Discussion}
\label{sec:discussion}

The systematic comparisons between T98 experimental data and detector simulations highlight how generator-level Final State Interactions (FSI), pion absorption, and secondary nucleon knockout shape observable distributions in a LArTPC:

\begin{itemize}
    \item \textbf{Vertex Energy Deposition and Nucleon Knockout}: The visible energy deposition near the annihilation vertex ($R < 5~\text{cm}$) receives a relatively large contribution from low-energy knockout protons, while heavy nuclear recoil fragments have a relatively smaller contribution due to recombination quenching in liquid argon. Consequently, generator predictions in this inner region serve as a direct probe of secondary nucleon emission models. The absence of secondary intranuclear cascade knockouts in Fritiof leads to an underestimation of local energy deposition. Conversely, INCL++ produces a high-energy proton tail that overestimates local energy loss. Empirically, GiBUU provides the closest match to the experimental energy profile across the full range. While this agreement can be qualitatively attributed to its continuous transport framework and post-processing potential barrier filter, the key practical conclusion is that GiBUU offers the most reliable phenomenological description of vertex-adjacent energy deposition in LAr.

    \item \textbf{Spatial Topology and Pion Absorption}: In outer radial shells ($R > 10~\text{cm}$), spatial hit cluster multiplicities serve as a sensitive benchmark for pion propagation and in-medium absorption. Fritiof predicts a higher primary pion multiplicity with minimal FSI, resulting in an overestimation of cluster counts at large radii. Both INCL++ and GiBUU account for in-medium pion absorption, bringing the spatial hit distributions into closer alignment with data. From a pragmatic standpoint, GiBUU achieves the best overall agreement across all radial shells, providing an accurate spatial template for LArTPC detector response modeling.
\end{itemize}

From a methodological standpoint, the data-driven detector simulation pipeline—incorporating the measured beam profile, Birks' law recombination, $X/(X+Y)$ spatial charge sharing, and TIGAr electronics transfer functions—proved highly effective. It enabled direct, quantitative Data-MC comparisons at the digitized waveform and cluster levels without requiring explicit 3D track reconstruction on complex multi-track star topologies.

These findings underscore the importance of advanced transport frameworks with full FSI when modeling low-energy antiproton interactions in heavy noble liquid detectors. The benchmarked response pipeline established in this work provides an essential foundational reference for future LArTPC-based physics programs—ranging from cosmic antimatter searches in experiments like GRAMS to rare-event probes of fundamental symmetries, such as neutron--antineutron ($n\text{--}\bar{n}$) oscillations.

\section{Conclusion}
\label{sec:conclusion}

In February 2025, the J-PARC T98 experiment acquired a high-statistics dataset of antiproton--argon ($\bar{p}\text{--}\text{Ar}$) interactions using a prototype single-phase LArTPC at the K1.8BR beamline. More than $10^5$ at-rest annihilation candidates were selected from data taken with a $0.7~\text{GeV}/c$ antiproton beam. Reference samples of cosmic-ray muons and beam particles were also collected to support detector calibration and response studies.

This work documents the experimental configuration, data selection, calibration procedures, and reconstruction methods used to characterize these events. A Geant4-based simulation framework was developed to incorporate the measured beam profiles, ionization charge production and recombination, charge transport and sharing, and electronics response. Cosmic-ray muons were used to establish the calorimetric calibration, which was checked against an independent sample of penetrating $1.0~\text{GeV}/c$ protons. Based on this comparison, a $3\%$ uncertainty was assigned to the calorimetric energy scale. Simulated waveforms were processed with the same reconstruction procedures as the experimental data, providing a common basis for comparisons of reconstructed observables.

As an initial application, simulations incorporating Fritiof, INCL++, and GiBUU were compared with data using the reconstructed total energy deposition, radial energy profiles, and spatial cluster multiplicities. These comparisons illustrate the use of the framework to characterize annihilation topologies and document differences between the data and the implemented models. The resulting distributions provide reference information for subsequent analyses, while detailed interpretations of the underlying nuclear reaction mechanisms remain a subject for future studies.

The dataset, calibration procedures, and simulation and reconstruction methods described here provide a common reference for future T98 physics analyses. Planned applications include particle identification in multi-track annihilation events, studies of charge-exchange reactions ($\bar{p} \to \bar{n}$), and extensions to light antinuclei such as $\bar{d}$ and $\bar{\text{He}}$. These resources also provide an experimental and computational basis for developing LArTPC-based antimatter searches, including GRAMS.

\section*{Acknowledgments}
We express our sincere gratitude to the J-PARC Hadron group for their excellent support in beam operation and facility management during the T98 experiment. This work was supported by JSPS KAKENHI Grant Numbers JP22H00133, JP23K22523, JP25K01017, and JP26H02036. The U.S. members of the team are partially supported by the NASA APRA GRAMS grant, No.22-APRA22-0128 (80NSSC23K1661).

\bibliographystyle{unsrt}
\bibliography{main}

@article{Chamberlain:1955ns,
    author = "Chamberlain, O. and Segre, E. and Wiegand, C. and Ypsilantis, T.",
    title = "{Observation of Anti-protons}",
    journal = "Phys. Rev.",
    volume = "100",
    pages = "947--950",
    year = "1955",
    doi = "10.1103/PhysRev.100.947"
}

@article{Klempt:2005pp,
    author = "Klempt, E. and Batty, C. and Richard, J. M.",
    title = "{The Antinucleon-nucleon interaction at low energy : Annihilation dynamics}",
    eprint = "hep-ex/0501020",
    archivePrefix = "arXiv",
    primaryClass = "hep-ex",
    journal = "Phys. Rept.",
    volume = "413",
    pages = "197--317",
    year = "2005",
    doi = "10.1016/j.physrep.2005.03.002"
}

@techreport{Rubbia:1977zz,
    author = "Rubbia, C.",
    title = "{The Liquid Argon Time Projection Chamber: A New Concept for Neutrino Detectors}",
    number = "CERN-EP-INT-77-08",
    institution = "CERN",
    year = "1977"
}

@article{Cholis:2020twh,
    author = "Cholis, I. and Linden, T. and Hooper, D.",
    title = "{Antideuterons and antihelium nuclei from annihilating dark matter}",
    eprint = "2001.08749",
    archivePrefix = "arXiv",
    primaryClass = "astro-ph.HE",
    journal = "Phys. Rev. D",
    volume = "102",
    number = "10",
    pages = "103019",
    year = "2020",
    doi = "10.1103/PhysRevD.102.103019"
}

@article{Mori:2002,
    author = "Mori, K. and Hailey, C. J. and Baltz, E. A. and Craig, W. W. and Kamionkowski, M. and Serber, W. T. and Ullio, P.",
    title = "{A novel antimatter detector based on X-ray deexcitation of exotic atoms}",
    eprint = "astro-ph/0109463",
    archivePrefix = "arXiv",
    primaryClass = "astro-ph",
    journal = "Astrophys. J.",
    volume = "566",
    pages = "604--616",
    year = "2002",
    doi = "10.1086/338253"
}

@article{Aramaki:2019bpi,
    author = "Aramaki, T. and Hansson Adrian, P. and Karagiorgi, G. and Odaka, H.",
    title = "{Dual MeV Gamma-Ray and Dark Matter Observatory - GRAMS Project}",
    eprint = "1901.03430",
    archivePrefix = "arXiv",
    primaryClass = "astro-ph.HE",
    journal = "Astropart. Phys.",
    volume = "114",
    pages = "107--114",
    year = "2020",
    doi = "10.1016/j.astropartphys.2019.07.002"
}

@article{Mohapatra:2009wp,
    author = "Mohapatra, R. N.",
    title = "{Neutron-Anti-Neutron Oscillation: Theory and Phenomenology}",
    eprint = "0902.0834",
    archivePrefix = "arXiv",
    primaryClass = "hep-ph",
    journal = "J. Phys. G",
    volume = "36",
    pages = "104006",
    year = "2009",
    doi = "10.1088/0954-3899/36/10/104006"
}

@article{DUNE:2020ypp,
    author = "Abi, B. and others",
    collaboration = "DUNE",
    title = "{Deep Underground Neutrino Experiment (DUNE), Far Detector Technical Design Report, Volume II: DUNE Physics}",
    eprint = "2002.03005",
    archivePrefix = "arXiv",
    primaryClass = "hep-ex",
    year = "2020",
    doi = "10.2172/1599307"
}

@article{LArIAT:2024lbx,
    author = "Basque, V. and others",
    collaboration = "LArIAT",
    title = "{First observation of antiproton annihilation at rest on argon in the LArIAT experiment}",
    eprint = "2409.13596",
    archivePrefix = "arXiv",
    primaryClass = "hep-ex",
    journal = "Phys. Rev. D",
    volume = "111",
    number = "9",
    pages = "094031",
    year = "2025",
    doi = "10.1103/PhysRevD.111.094031"
}

@article{Agari:2012_K18BR,
    author = "Agari, K. and others",
    title = "{The K1.8BR beamline at the J-PARC Hadron Hall}",
    journal = "Prog. Theor. Exp. Phys.",
    volume = "2012",
    pages = "02B008",
    year = "2012",
    doi = "10.1093/ptep/pts039"
}

@article{Agostinelli:2003_Geant4,
    author = "Agostinelli, S. and others",
    collaboration = "GEANT4",
    title = "{GEANT4--a simulation toolkit}",
    journal = "Nucl. Instrum. Meth. A",
    volume = "506",
    pages = "250--303",
    year = "2003",
    doi = "10.1016/S0168-9002(03)01368-8"
}

@article{Andersson:1987_FTF,
    author = "Andersson, B. and Gustafson, G. and Nilsson-Almqvist, B.",
    title = "{A Model for Low pT Hadronic Reactions, with Generalizations to Hadron - Nucleus and Nucleus-Nucleus Collisions}",
    journal = "Nucl. Phys. B",
    volume = "281",
    pages = "289--309",
    year = "1987",
    doi = "10.1016/0550-3213(87)90257-4"
}

@article{Mancusi:2014_INCL,
  author        = {Mancusi, Davide and Boudard, Alain and Cugnon, Joseph and David, Jean-Christophe and Leray, Sylvie and Rodriguez-Sanchez, David},
  title         = {Extension of the Li{\`e}ge intranuclear-cascade model {INCL++} to reactions induced by light ions up to cluster spallation},
  journal       = {Phys. Rev. C},
  volume        = {90},
  pages         = {064602},
  year          = {2014},
  doi           = {10.1103/PhysRevC.90.064602}
}

@phdthesis{Zharenov:2023_INCL_pbar,
  author        = {Zharenov, Denis},
  title         = {Interaction of antiprotons with nuclear matter within the {INCL} model},
  school        = {Universit{\'e} Paris-Saclay},
  year          = {2023},
  type          = {{PhD} thesis},
  url           = {https://theses.hal.science/tel-04511526}
}

@article{Buss:2012_GiBUU,
    author = "Buss, O. and Gaitanos, T. and Gallmeister, K. and van Hees, H. and Kaskulov, M. and Lalakulich, O. and Larionov, A. and Leitner, T. and Mosel, U. and Alvarez-Ruso, L. and others",
    title = "{Transport-theoretical description of nuclear reactions}",
    eprint = "1106.1344",
    archivePrefix = "arXiv",
    primaryClass = "hep-ph",
    journal = "Phys. Rept.",
    volume = "512",
    pages = "1--124",
    year = "2012",
    doi = "10.1016/j.physrep.2011.12.001"
}

@techreport{Sanford:1967_SW,
    author = "Sanford, J. R. and Wang, C. L.",
    title = "{Empirical formulas for particle production in p-Be collisions between 10 and 35 BeV}",
    number = "BNL-11299",
    institution = "Brookhaven National Laboratory",
    year = "1967"
}

@article{Ieiri:2008zzc,
    author = "Ieiri, Masaharu and others",
    title = "{Electrostatic separator for K1.8 beam line at J-PARC}",
    doi = "10.1016/j.nimb.2008.05.040",
    journal = "Nucl. Instrum. Meth. B",
    volume = "266",
    pages = "4205--4208",
    year = "2008"
}

@article{Tanaka:2020_LTARS,
    author = "Tanaka, M. and others",
    title = "{Development of front-end ASIC for liquid argon time projection chamber}",
    journal = "JINST",
    volume = "15",
    number = "09",
    pages = "T09009",
    year = "2020",
    doi = "10.1088/1748-0221/15/09/T09009"
}

@article{Aoyama:2021xzx,
    author = "Aoyama, K. and Tanaka, M. and Kimura, M. and Yorita, K.",
    title = "{Development of a liquid argon detector with high light collection efficiency using tetraphenyl butadiene and a silicon photomultiplier array}",
    eprint = "2105.04285",
    archivePrefix = "arXiv",
    primaryClass = "physics.ins-det",
    journal = "PTEP",
    volume = "2022",
    number = "4",
    pages = "043H01",
    year = "2022",
    doi = "10.1093/ptep/ptac064"
}

@patent{Hough:1962,
    author = "Hough, P. V. C.",
    title = "{Method and means for recognizing complex patterns}",
    number = "US3069654A",
    year = "1962"
}

@article{Duda:1972,
    author = "Duda, R. O. and Hart, P. E.",
    title = "{Use of the Hough Transformation to Detect Lines and Curves in Pictures}",
    journal = "Commun. ACM",
    volume = "15",
    number = "1",
    pages = "11--15",
    year = "1972",
    doi = "10.1145/361237.361242"
}

@article{Miyajima:1974_Wion,
    author = "Miyajima, M. and Takahashi, T. and Konno, S. and Hamada, T. and Kubota, S. and Shibamura, H. and Doke, T.",
    title = "{Average energy expended per ion pair in liquid argon}",
    journal = "Phys. Rev. A",
    volume = "9",
    pages = "1438--1443",
    year = "1974",
    doi = "10.1103/PhysRevA.9.1438"
}

@article{ICARUS:2004koz,
    author = "Amoruso, S. and others",
    collaboration = "ICARUS",
    title = "{Study of electron recombination in liquid argon with the ICARUS TPC}",
    journal = "Nucl. Instrum. Meth. A",
    volume = "523",
    pages = "275--286",
    year = "2004",
    doi = "10.1016/j.nima.2003.11.423"
}

@article{Li:2016_diffusion,
    author = "Li, Y. and others",
    title = "{Measurement of longitudinal electron diffusion coefficient in liquid argon}",
    eprint = "1508.07059",
    archivePrefix = "arXiv",
    primaryClass = "physics.ins-det",
    journal = "Nucl. Instrum. Meth. A",
    volume = "816",
    pages = "160--170",
    year = "2016",
    doi = "10.1016/j.nima.2016.01.094"
}

@article{Shibamura:1979_diffusion,
    author = "Shibamura, E. and Takahashi, T. and Kubota, S. and Doke, T.",
    title = "{Ratio of diffusion coefficient to mobility for electrons in liquid argon}",
    journal = "Phys. Rev. A",
    volume = "20",
    pages = "2547--2554",
    year = "1979",
    doi = "10.1103/PhysRevA.20.2547"
}

\end{document}